\documentclass{aa}
\usepackage{natbib}
\usepackage{color}
\usepackage[varg]{txfonts}
\usepackage{graphicx}

\begin{document}

\title{Transient evolution of C-type shocks in dusty regions of varying density}

\author{ I.~Ashmore\inst{1}, S. Van Loo\inst{1}, 
P.~Caselli\inst{1}, S.~A.~E.~G.~Falle\inst{2}, \and T.~W.~Hartquist\inst{1}}
\authorrunning{Ashmore et al.}

\offprints{I. Ashmore \email{phy9ia@leeds.ac.uk}}

\institute{School of Physics and Astronomy, University of Leeds,
        Leeds LS2 9JT, UK \and School of Mathematics, University of Leeds,
        Leeds LS2 9JT, UK}

\date{Received date; accepted date}

\abstract
        {Outflows of young stars drive shocks into dusty, 
	molecular regions. Most models of such shocks are restricted by 
	the assumptions that they are steady and propagating in directions
	perpendicular to the magnetic fields. However, the media through 
	which shocks propagate are inhomogeneous and shocks are not
	steady. Furthermore, only a small fraction of shocks are 
	nearly perpendicular.}
        {We identify features that develop when 
	a shock encounters a density inhomogeneity and 
	ascertain if any part of the precursor region of a non-steady 
	multifluid shock ever behaves in a quasi-steady fashion. 
	If it does, some time-dependent shocks may be modelled approximately
	without solving the time-dependent hydromagnetic equations.}
	{We use the code employed previously to produce the first 
	time-dependent simulations of fast-mode oblique C-type shocks
	including a self-consistent calculation of the thermal and 
	ionisation balances and a fluid treatment of grains.}
        {Simulations were made for initially steady oblique C-type shocks,
	each of which encounters one of three types of density inhomogeneities.
	For a semi-finite inhomogeneity with a density larger than 
	the surrounding medium's, a transmitted shock evolves from being of 
	J-type to a steady C-type shock on a timescale comparable to the 
	ion-flow time through it. A sufficiently upstream part of 
	the precursor of an evolving J-type shock is quasi-steady. The 
	ion-flow timescale is also relevant for the evolution of a 
	shock moving into a region of decreasing density. The models 
	for shocks propagating into regions in which the density increases 
	and then decreases to its initial value cannot be
	entirely described in terms of the results obtained for 
	monotonically increasing and decreasing densities.}
	{We present the first time-dependent simulations of dusty C-type 
	shocks interacting with density perturbations. We studied the 
	transient evolution of the shock structure and find that the initial 
	interaction always produces a transition to a J-type shock. 
	Furthermore, the long-term evolution back to a C-type shock cannot 
	always be approximated by quasi-steady models.}
	
\keywords{MHD -- shock waves -- ISM: dust -- ISM: jets and outflows}

\maketitle

\section{Introduction}

Jets and winds associated with low and high mass 
proto-stellar objects interact with the molecular clouds surrounding them 
\citep[e.g.][]{Aal07}. Shocks are driven into the 
clouds sweeping up cloud material and producing large scale 
(0.1 - 1~pc) molecular outflows with velocities of 10-100 km s$^{-1}$ 
\citep[e.g.][]{BT99,RB01,SG09}. These outflows are often bipolar and are 
extremely common in young, low-mass stars \citep{BL83}, as well as high-mass
stars \citep{C05,S05,LS09}.

As the fractional ionisation in molecular clouds is low 
\citep[$\chi < 10^{-6}$;][and reference therein]{D06}, 
the neutral gas and  magnetic field are weakly coupled. This coupling is 
mediated via ion-neutral and grain-neutral collisions, as the magnetic 
field forces the 
charged particles to move through the bulk of neutral particles. 
When a shock forms, the ion-neutral collisions accelerate, compress and 
heat the neutral gas upstream of the shock, forming a magnetic precursor. 
If the shock velocity is low and the postshock cooling efficient, the shock 
becomes continuous in all fluids and is then referred to as a C-type shock 
\citep[][]{M71,D80}. Observational studies of the velocity widths of 
SiO emission suggest that most shocks near low mass proto-stellar objects 
are C-type \citep[e.g.][]{MPal92}. 

Previous studies of C-type shocks in dusty plasmas have focused on the 
shock structure under steady state conditions 
\citep[e.g.][]{DRD83, PHH90, PH94, W98, G08}. However, SiO observations 
suggest that proto-stellar jets and winds interact with clumpy structures 
along their propagation axis \citep{Mal92, Lal98, Jetal04}. Therefore, 
the steady state assumption must be relaxed. \cite{CR02} were the 
first\footnote{Time-dependent multifluid shock studies 
prior to \cite{CR02},  e.g. those of \cite{T94, SM97, S97} and \cite{CPDFF98}, 
did not include grain dynamics.} 
to use a time-dependent multifluid magnetohydrodynamics (MHD) code,
including dust grains dynamics, to study fast-mode C-type shocks. 
However, their study is limited to perpendicular 
shocks. The multifluid approach developed by \cite{F03} overcomes this 
restriction.  Using the Falle method, \citet[][hereafter Paper~I]{vLal09} 
produced the first time-dependent simulations of fast-mode, 
oblique C-type shocks that evolve to steady state  and include a 
selfconsistent calculation of the thermal 
and ionisation balances and a fluid treatment of grain dynamics.

While the time-dependent simulations of Paper I still focus on steady state
shock structures, we now extend this work by studying the temporal evolution of 
fast-mode C-type shocks interacting with regions of inhomogeneous density. 
In Sect.~\ref{sect:model} we describe the numerical model with
the initial conditions. We then apply the code to an oblique 
shock propagating into a region of varying density (Sect.~\ref{sec:results})
and, in Sect.~\ref{sect:conc}, we discuss these results and give our 
conclusions. 

\section{The model}\label{sect:model}
\subsection{Numerical code}\label{subsec:numcod}
The numerical code that we use in this paper follows the dynamics of 
a multifluid plasma consisting of neutrals, ions, electrons and N dust grain 
species. Here we only consider a single fluid of spherical, large grains 
with a radius and mass of $a_g = 0.4 \mu$m and 8.04 $\times$ 10$^{-13}$g. 
(The density of the grains is assumed to be 3 g cm$^{-3}$.) We have included
relevant mass transfer processes, such as electron recombination with Mg$^+$
and dissociative recombination with  HCO$^+$, and radiative cooling
by O, CO, H$_2$ and H$_2$O. Also, the mass and charge transfer from ions and 
electrons to the dust grains is taken into account to  calculate the average 
grain charge. This way we calculate self-consistently the thermal and 
ionisation balances including appropriate microphysics.

The details of the numerical scheme, which is based on 
\cite{F03}, are given in Paper I. As in Paper~I, we advance the scheme 
explicitly, even though this implies a restriction on the stable time-step, 
i.e. \cite{F03}
\[
   \Delta t \leq \frac{r_{ad}}{(r^2_H + r^2_{ad})} \Delta x^2,
\]
where $r_{ad}$ is the ambipolar resistivity, $r_H$ the Hall resistivity and 
$\Delta x$ the cell spacing. Thus, 
the time step becomes small for a high numerical resolution, or when the 
Hall resistivity becomes large compared to the ambipolar resistivity. In our
simulations the Hall resistivity is only ever comparable to the ambipolar 
resistivity implying that the stable time-step is not severely restricted.

The adopted chemistry is based on the network given in \cite{PHH90},
and we used the rate coefficients that they did. The code
allows the option of incorporating as many advected scalar equations with 
source terms (i. e. rate equations for fractional abundances) as necessary
and solving them with the same methods as those employed to solve the
fluid equations. We used rate equations and charge neutrality to
obtain the fractional abundances of ions and electrons and the average
grain charge. We assumed that cosmic ray induced ionisation leads
immediately to the production of heavy molecular ions (e.g. HCO$^+$
and H$_3$O$^+$) which involves the justifiable neglect of $\rm {H_3}^+$
recombination. The branching ratio between HCO$^+$ and H$_3$O$^+$
formation depends on the abundances of simple neutral species like
O and CO and the rate coefficients of the reaction rates
of $\rm {H_3}^+$ with them. Charge transfer with metals (e.g. Mg) creates
metallic ions. All gas phase ions recombine with electrons and on grains. 
Unlike \cite{PHH90}, we did not include the possibility that
Mg$^+$ reacts with H$_2$; it does so only in regions with sufficiently 
high temperatures that the grain charge is high enough and the electron
fractional abundance low enough that recombination is dominated
by recombination with grains leading to a negligible difference between 
the rates at which Mg$^+$ and MgH$^+$ are removed. While the abundances
of ions and electrons and grain charge can be studied easily with rate 
equations, the conversion of O to H$_2$O is more problematic as
the relevant reaction rates increase by many orders of magnitude
as shock gas is heated from several hundred to a thousand degrees.
For an H$_2$ number density of $10^5$ cm$^{-3}$, the
timescale for the removal of O in reactions with H$_2$ is
1.4 $\times 10^{12}$ s, 1.4 $\times 10^9$ s, and 6.7 $\times 10^7$ s at
300K, 600K, and 1000K, respectively. The corresponding removal
timescale of OH in reactions with H$_2$ are 1.5 $\times 10^9$ s,
3.2 $\times 10^7$ s, and 4.2 $\times 10^6$ s. We could use the code
to deal with the O, OH, and H$_2$O chemistry even at temperatures
in excess of $10^3$K if we placed upper limits on the reaction
rates; this would still give oxygen chemistry that would be close to
the appropriate high temperature equilibrium. However, here we
simply assumed that almost all oxygen not in CO is in O until
T = 500K. After it reached 500K almost all oxygen not in CO was taken to 
be in H$_2$O. It will remain mostly in H$_2$O for very long times
even after the postshock gas has cooled. This approach led to neutral
temperatures in the 300-500K range that are somewhat higher and 
in the 500-700K range that are somewhat lower than a more thorough 
treatment would have. However, this results in a negligible difference
in the shock behaviour which is much more strongly influenced
by the ionisation structure, which in turn is regulated in this regime
primarily by cosmic ray induced ionization and recombination
on grains, and grain dynamics than the details of the neutral
gas temperature structure.

\subsection{Initial conditions}\label{subsec:initcond}
Unlike the simulations described in Paper I, for which 
the initial conditions are J-type shocks, 
each of our simulations starts from a steady C-type shock propagating 
through a medium with upstream hydrogen nuclei number density of 
either $n_{\rm H} = 10^4 {\rm cm^{-3}}$ or 
$n_{\rm H} = 10^5 {\rm cm^{-3}}$. The shock moves parallel to 
the $x$-axis with a velocity $V_s$ of 25 km s$^{-1}$. 
For both cases the initial 
upstream fluid temperatures are 8.4K and the upstream magnetic field strengths 
are B = 10$^{-4}$G. The magnetic field lies in the $x,y$-plane with the angle 
between the shock normal and the field being 45$^\circ$. 

As we are interested in the interaction of a C-type shock with an 
inhomogeneous medium, we introduce an upstream density perturbation.
The shape of the perturbation is chosen so that it reasonably 
reproduces all or part of the density profile of observed clumps or 
cores \citep[e.g.][]{Tal04}. Each perturbation is described solely by its 
width and the density contrast (i.e. a multiple of initial upstream density). 
Unless otherwise stated, the width and density contrast are 10$^{17}$ cm and 
10 respectively. 

\subsection{Computational considerations}\label{subsec:compdet}
\subsubsection{Computational grid}
 Some care must be given to the size of the computational grid for models 
with varying upstream density. This is because a change in the upstream 
density alters the shock width. For all our simulations, the constraint
\[
   \chi < 3 \times 10^{-9} \left(\frac{V_s}{\rm km\ s^{-1}}\right)
\]
is met (with $\chi$ the fractional ionisation). The shock thickness is 
then determined by grain-neutral rather 
than ion-neutral drag (see Paper I) and given approximately by
\begin{equation}\label{eq:shockwidths}
   L = 2.1 \times 10^{24} \chi \left(\frac{V_s}{\rm km\ s^{-1}}\right)^{-1} 
	\left(\frac{v_A}{\rm km\ s^{-1}}\right) 
	\left(\frac{10^{-2} {\rm cm}^{-3}}{n_i} \right) {\rm cm},
\end{equation}
where $v_{A}$ is the Alfv\'en speed and n$_{i}$ the ion number density. 
The shock width in our simulations varies by an order of a magnitude. 
We need to ensure that the numerical domain is large enough to contain 
the largest shock width but has enough cells to resolve structures on the 
finest scale, i.e the shock width at the highest density. Therefore, we 
estimate the shock thickness on both densities using 
Eq.~\ref{eq:shockwidths}. The numerical domain is then set to
$-10 L < x < 10 L$ with $L$ the shock width at the lowest density. As the 
shock width
at the highest density is roughly an order of magnitude smaller and we want 
to resolve it with approximately 10 cells across the shock structure, 
we adopt a uniform grid of 1200 cells for the numerical domain.
We impose a fixed inflow at the upstream boundary and a free-flow boundary 
condition at the downstream side.

\subsubsection{Varying shock speed}
When a shock moves through a region of varying density, its propagation speed 
changes. As we follow the evolution in the initial shock frame, the 
new shock structures (see Sect.~\ref{sec:results}) will eventually 
move off the grid. 
In order to follow their evolution properly, we adjust the velocity of 
the initial shock frame with a Lorentz transformation along the $x$-axis.

\subsubsection{Charged fluid inertia}
Although our initial conditions are given by a C-type shock, 
the interaction with a density inhomogeneity produces a neutral subshock
(see Sect.~\ref{sec:results}).
Within the subshock, our assumption that the charged fluid inertia is 
negligible no longer holds. However, the neutral subshock structure can be 
accurately modelled with our numerical model \citep[see][]{F03}. Furthermore,
the inertial phase is expected to be shortlived compared to the timescales of 
the shock-clump interaction and the subsequent approach to steady state. 
We can expect that the evolution of the shock shortly after the interaction
is modified by charged fluid inertia, but that the changes are modest. 
  
\begin{figure*}
\begin{center}
\includegraphics[width = 8.4 cm]{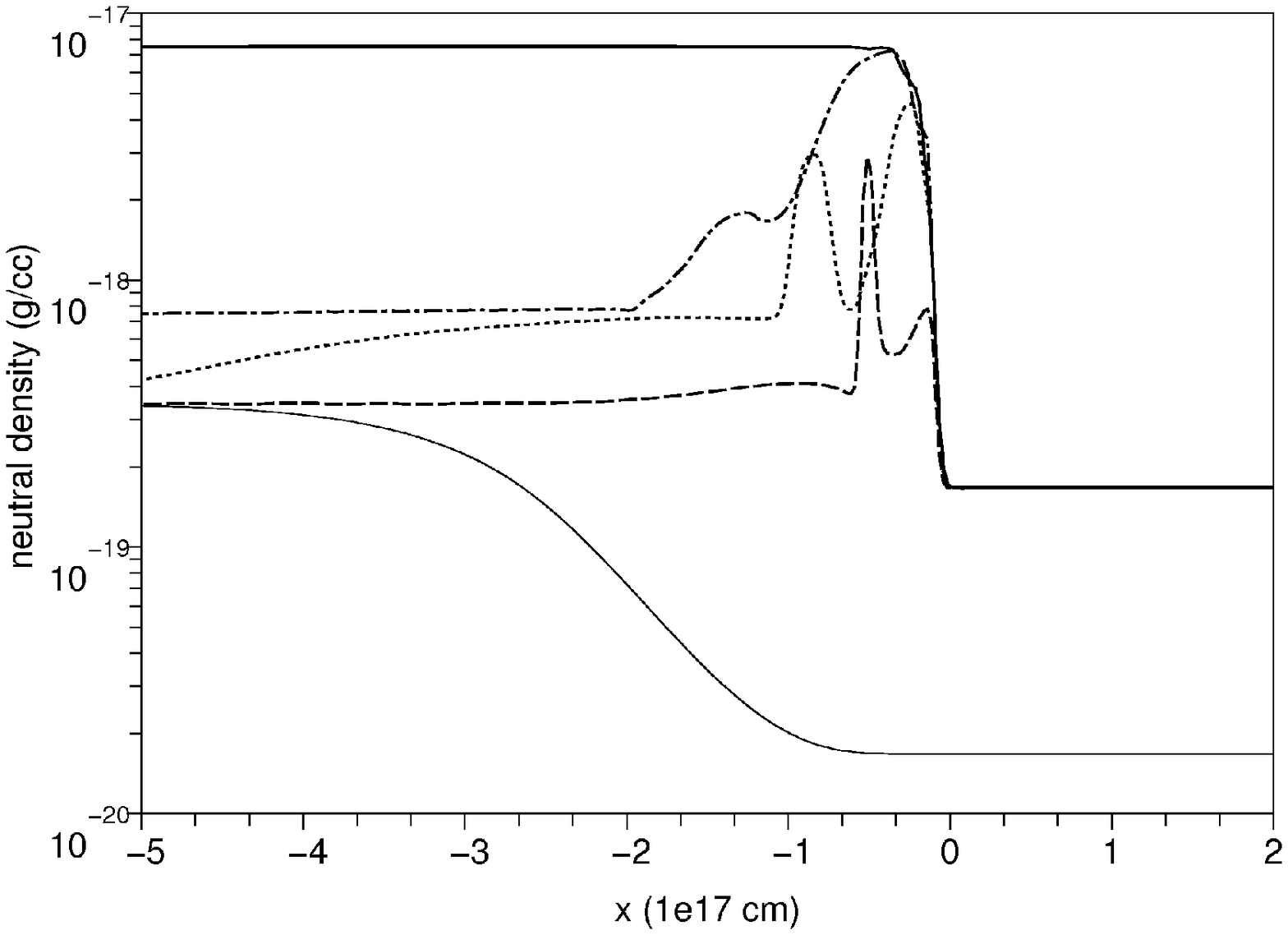}
\includegraphics[width = 8.4 cm]{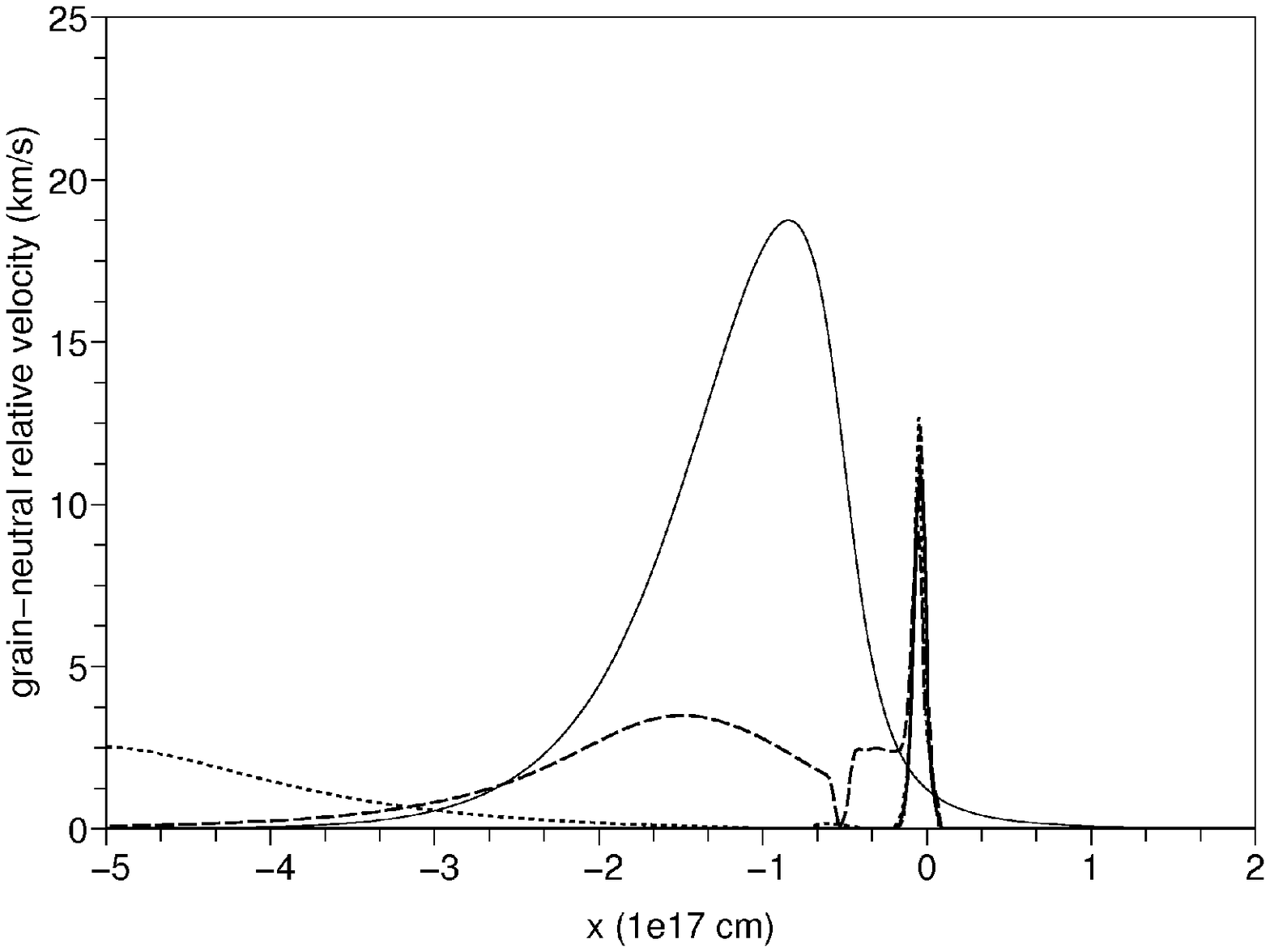}
\caption{Evolution of the shock structure in the shock frame as a 
25~km\ s$^{-1}$ C-type shock interacts with 
a inhomogeneity of increasing density (from $n_{\rm H} = 10^4 {\rm cm^{-3}}$ 
to $n_{\rm H} = 10^5 {\rm cm^{-3}}$). The left panel shows the neutral 
density and the right panel the grain-neutral relative speed.
The profile for the initial C-type shock is given by the thin solid line, while  
the dashed line shows the shock structure 9.75 $\times 10^3$~yr after the shock 
first interacts with the perturbation, the dotted line after 
$2.56\times 10^4$ yr, the dash-dotted line after $6.37\times 10^{4}$ yr, 
and the thick solid line the final steady C-type shock structure 
(reached after $7.92 \times 10^4$~yr).}
\label{fig1}
\end{center}
\end{figure*}

\section{Results}
\label{sec:results}

Several aspects of the shock interaction with an inhomogeneity are 
consistent for all our simulations. We discuss these first before
commenting on specific models.

In each of our models the initial shock width is larger than 
or comparable to the 
length scale of the perturbation. The interaction with the perturbation
then leads to a transmitted wave-reflected wave pair separated by a contact
discontinuity. Note that, if the perturbation length scale were large 
compared to the shock width, there would not be a reflected wave.  
The relative strength of the transmitted and reflected 
components is determined by both the initial shock speed and 
the amplitude of the perturbation.

The transmitted shock is initially a J-type shock with a precursor and a 
subshock. The upstream conditions change faster than information can propagate
across the front of the subshock. After some time, collisions between
ions and neutrals generate a neutral precursor which then evolves into a C-type 
shock. The transmitted C-type shock is narrower than the initial shock if 
the upstream density increases and broader if the upstream density decreases.

While the transmitted wave is always a shock, the reflected wave which 
propagates into the post-shock flow can either be a rarefaction wave 
or a shock. As the reflected wave connects the far downstream flow with 
the downstream flow of the transmitted shock, it is clear that a rarefaction
forms when the post-shock density of the transmitted shock is lower than 
the far downstream density and a shock when it is greater. Similarly
to the transmitted shock, the reflected shock evolves from a J-type shock
into a C-type shock.

In this paper we focus primarily on the transmitted shocks. We do not 
follow the long term evolution of the reflected waves. The reflected component 
thus eventually moves off the grid at the downstream boundary. 
However, numerical artefacts arise when shocks are reflected from 
free-flow boundaries \citep[e.g.][]{H79}. Although we can use 
non-reflecting boundary conditions to overcome this problem, we prefer to simply
remove the reflected wave from the computational domain. This can be 
done because the reflected wave and transmitted shock are separated by 
a contact discontinuity.

\begin{figure*}
\begin{center}
\includegraphics[width = 7.6 cm]{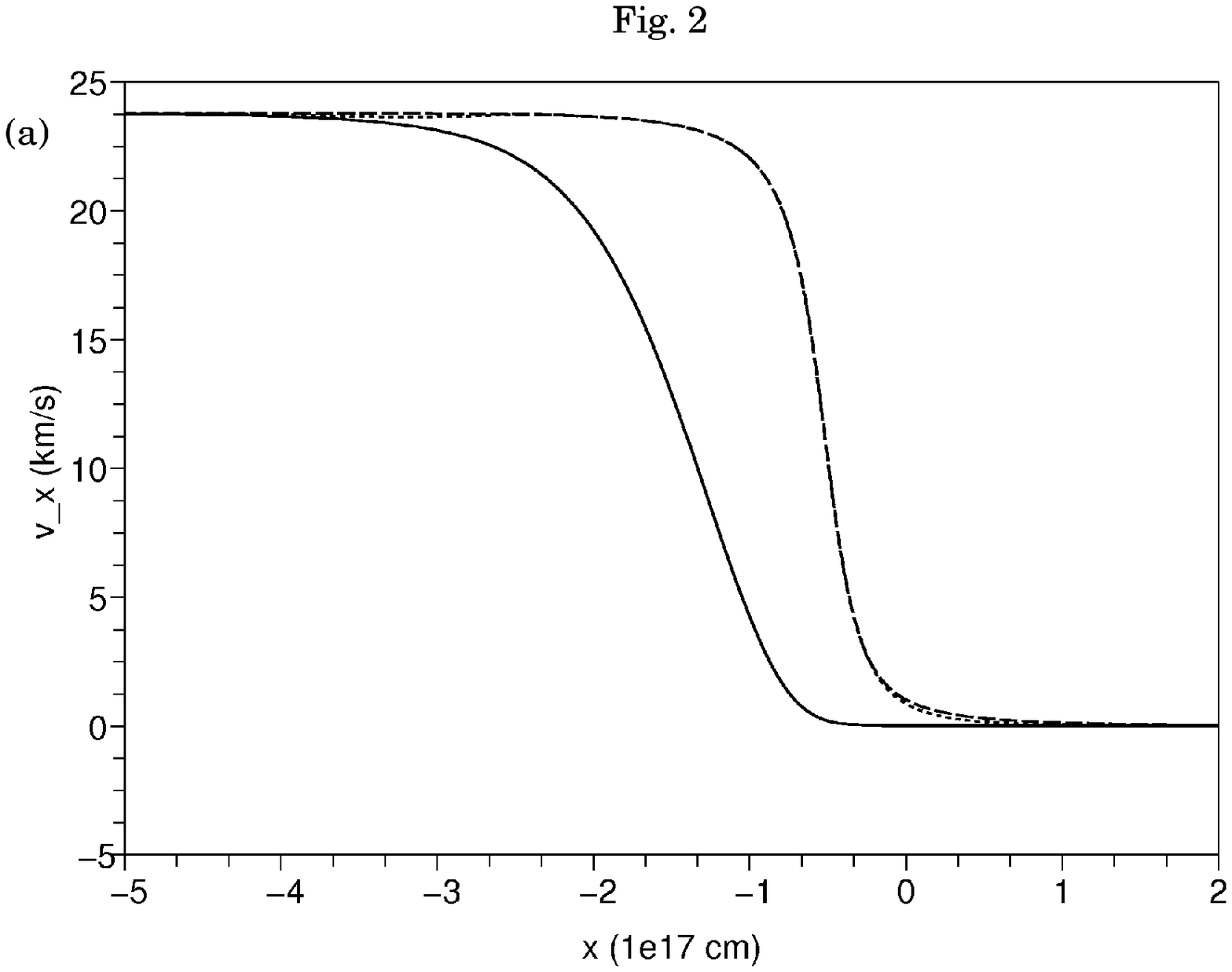}
\includegraphics[width = 7.6 cm]{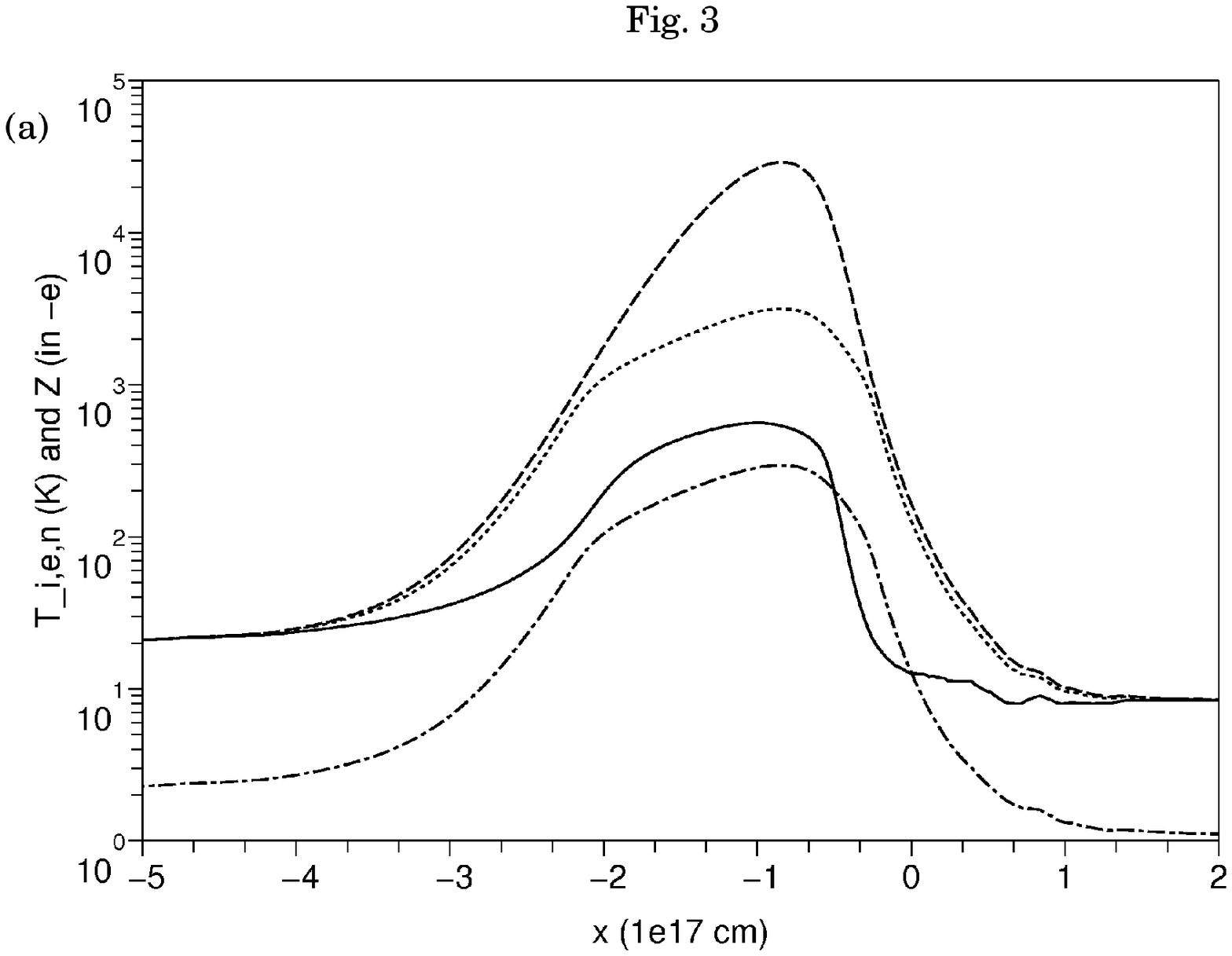}
\includegraphics[width = 7.6 cm]{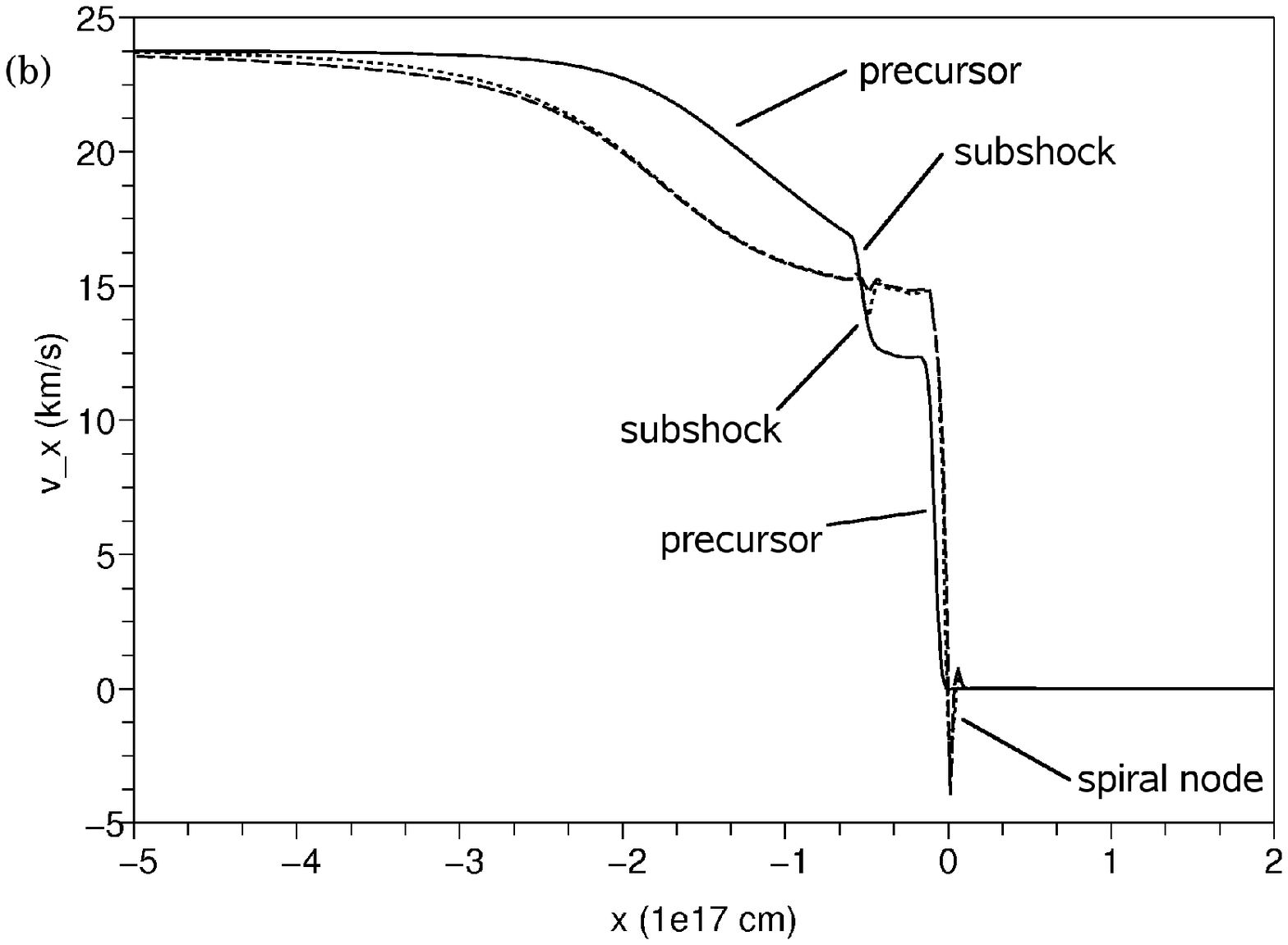}
\includegraphics[width = 7.6 cm]{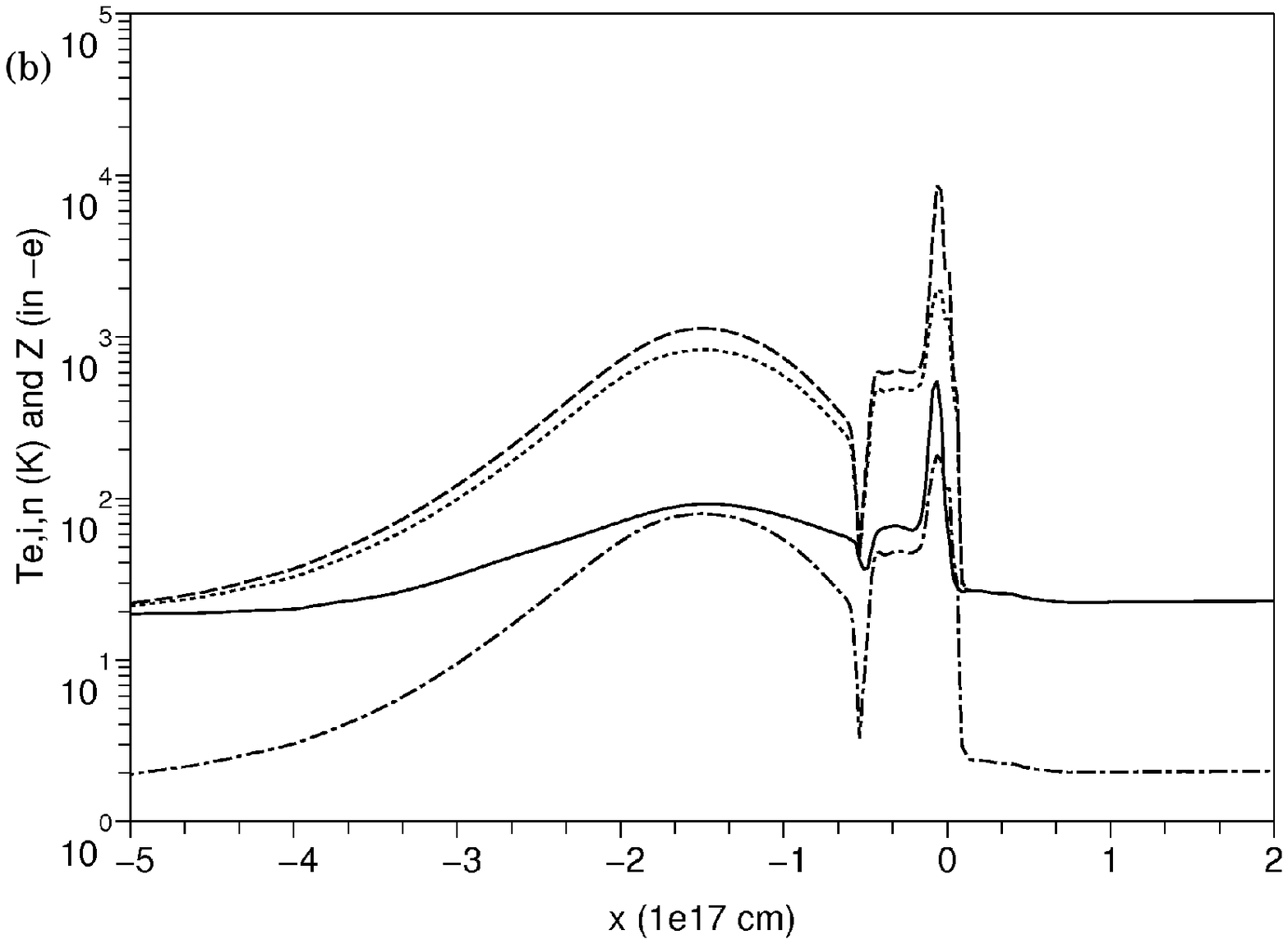}
\includegraphics[width = 7.6 cm]{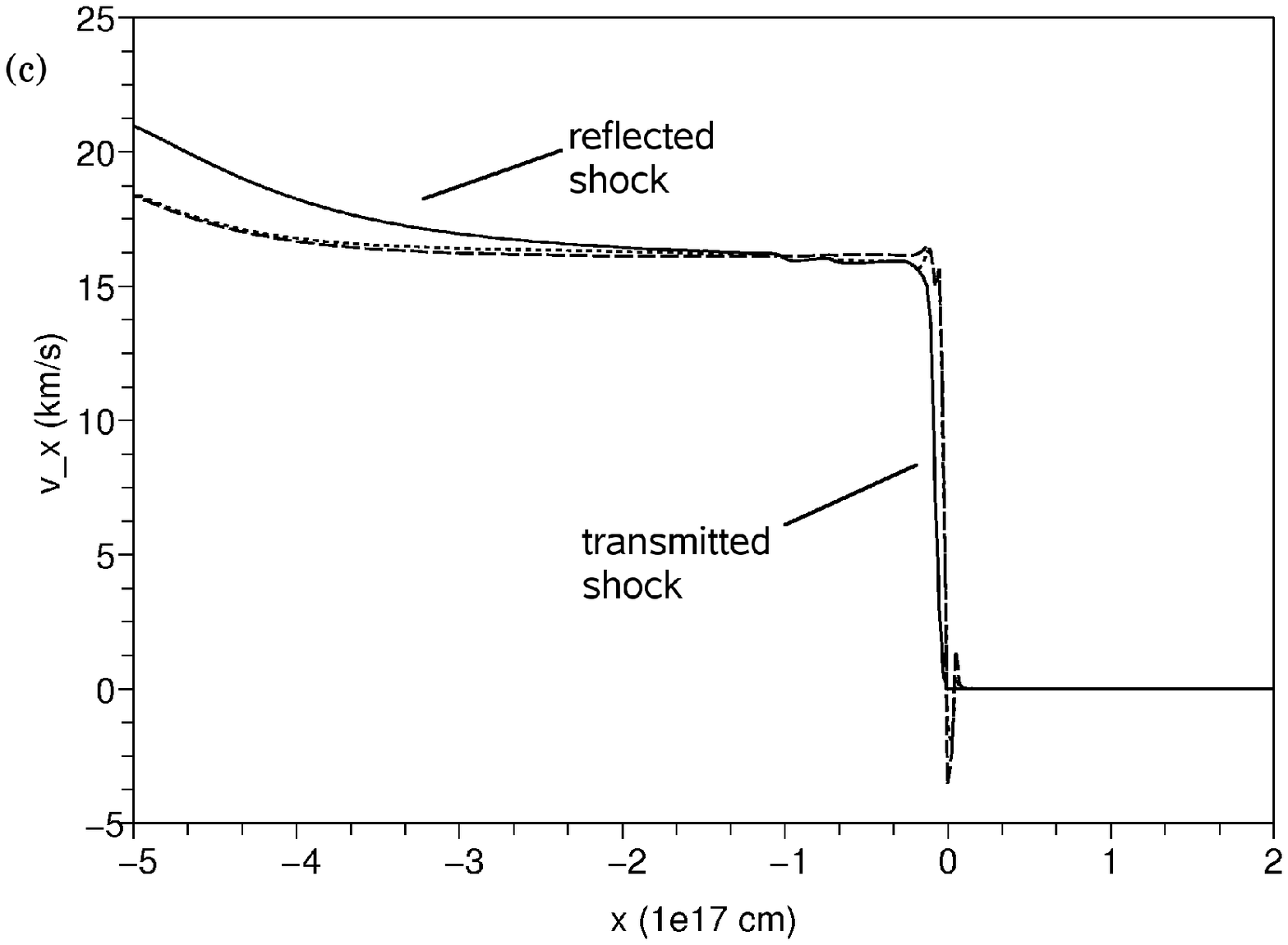}
\includegraphics[width = 7.6 cm]{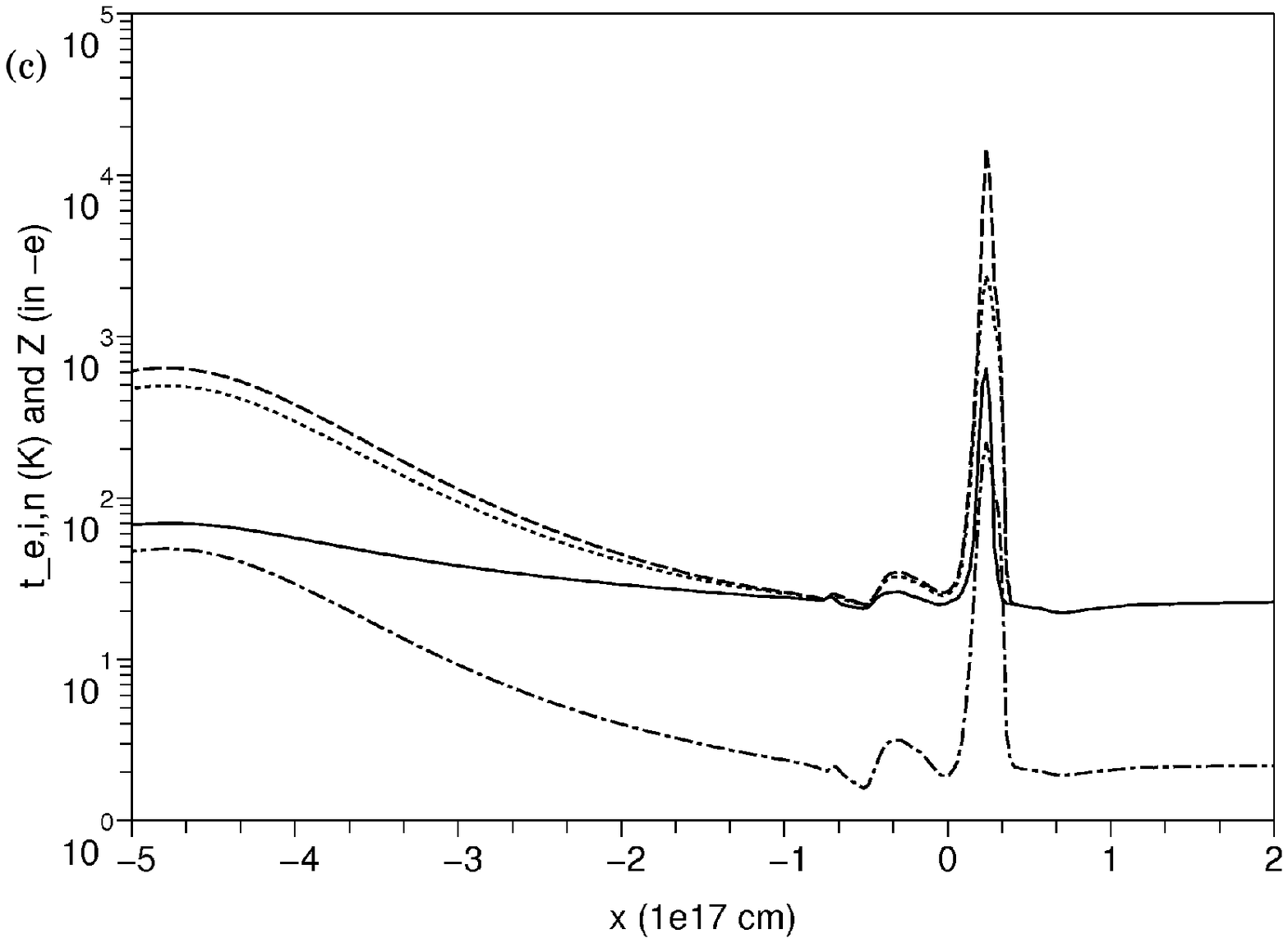}
\includegraphics[width = 7.6 cm]{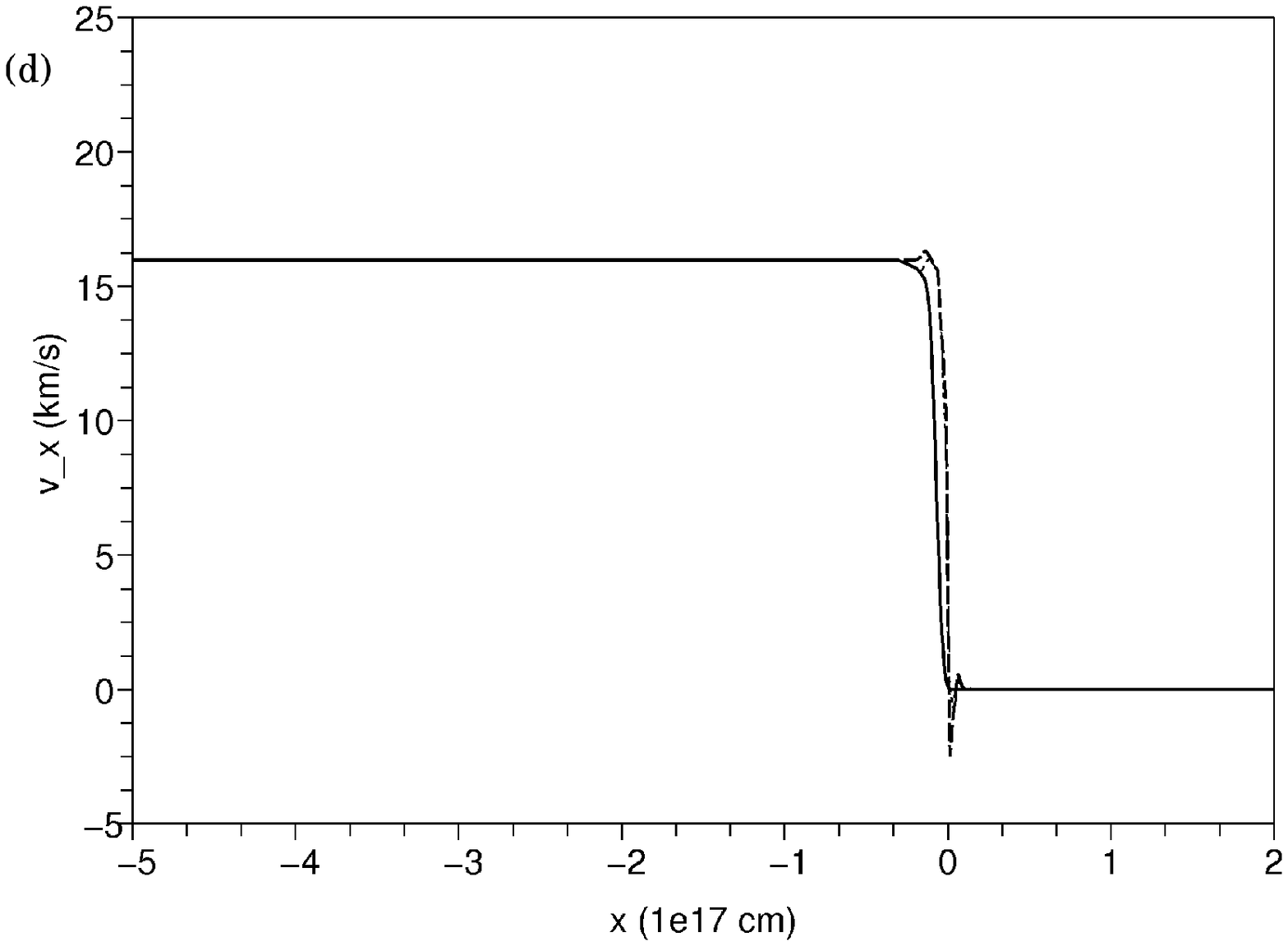}
\includegraphics[width = 7.6 cm]{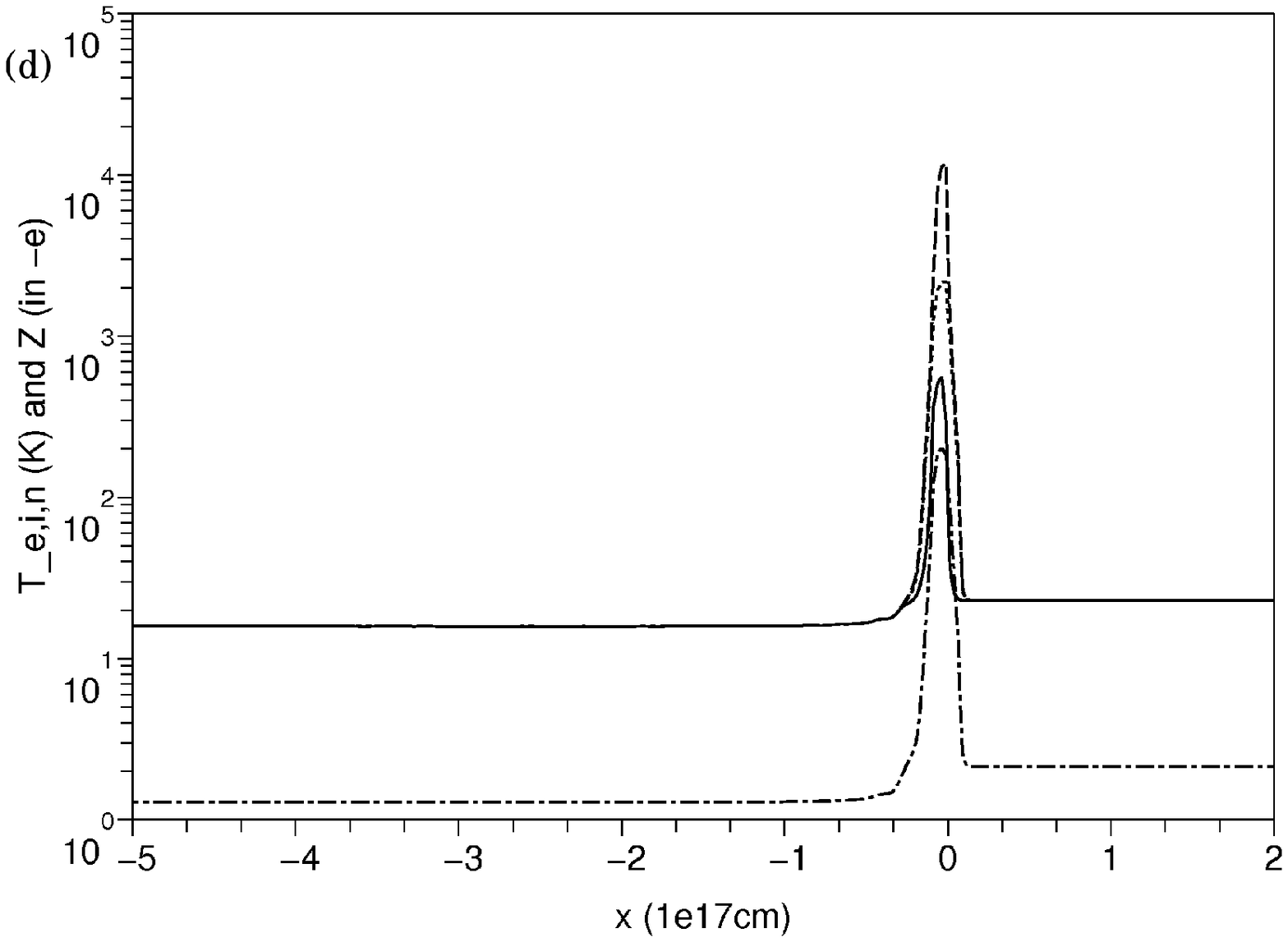}
\caption{$x$ component of the neutrals (solid), ions and electrons (dashed) and 
the grains (dotted) velocity in the  upstream frame at different times for
a 25 km\ s$^{-1}$ C-type shock propagating in a higher density region 
(from $n_{\rm H} = 10^4 {\rm cm^{-3}}$ to $n_{\rm H} = 10^5 {\rm cm^{-3}}$).
(a) shows the the initial C-type shock, (b) the  the shock structure
9.75 $\times 10^3$~yr after the first interaction of the shock with
the inhomogeneity, (c) after  $2.56\times 10^4$ yr and (d) the final
steady C-type shock (after $7.92 \times 10^4$~yr).}
\label{fig2}
\caption{Ion (dashed), electron (dotted) and neutral
temperature (solid) and the average grain charge (dash-dotted) in
electron charge units for the same shock model as in Fig.~\ref{fig2}
at the same times.}
\label{fig3}
\end{center}
\end{figure*}

\subsection{Increasing density}\label{subsec:constlow-high}
We first follow  the evolution of a C-type shock in the molecular outflow 
from a proto-stellar object moving into a denser region. 
We consider a 25 km\ s$^{-1}$ shock propagating 
from a region of $n_{\rm H} = 10^4 {\rm cm^{-3}}$ into a higher density
plasma. The initial shock is a strong fast-mode shock with an 
Alfv\'enic Mach number of $\approx 11.5$.

When the C-type shock encounters the density perturbation, two J-type shocks, 
a reflected and a transmitted one, form within $3 \times 10^3$ yr. The neutral 
subshocks can be easily seen in both the velocity and temperature plots of 
Figs.~\ref{fig1}, \ref{fig2} and \ref{fig3}, i.e. the discontinuous jump in the 
neutral velocity along the $x$-axis and the drastic cooling of the 
different fluids. The temperature plots also shows that the transmitted shock 
is considerably stronger than the reflected one, as the maximum temperature
in the transmitted shock is roughly 5 times higher. The temperatures of 
both the transmitted and reflected shocks decrease at later times. 
This is due to a reduced relative velocity between the neutrals and the 
charged particles as the energy transfer by collisions heats the gas.  
The lower temperatures and relative velocities suggest that the higher 
upstream density regime is less effective for grain sputtering. This agrees 
with the results of \cite{Mal00} who reported that the elemental fraction 
of Si sputtered from olivine increased with shock velocity, but at 
a given shock velocity, dropped as the pre-shock density increased.  

\begin{figure*}
\begin{center}
\includegraphics[width = 8.4 cm]{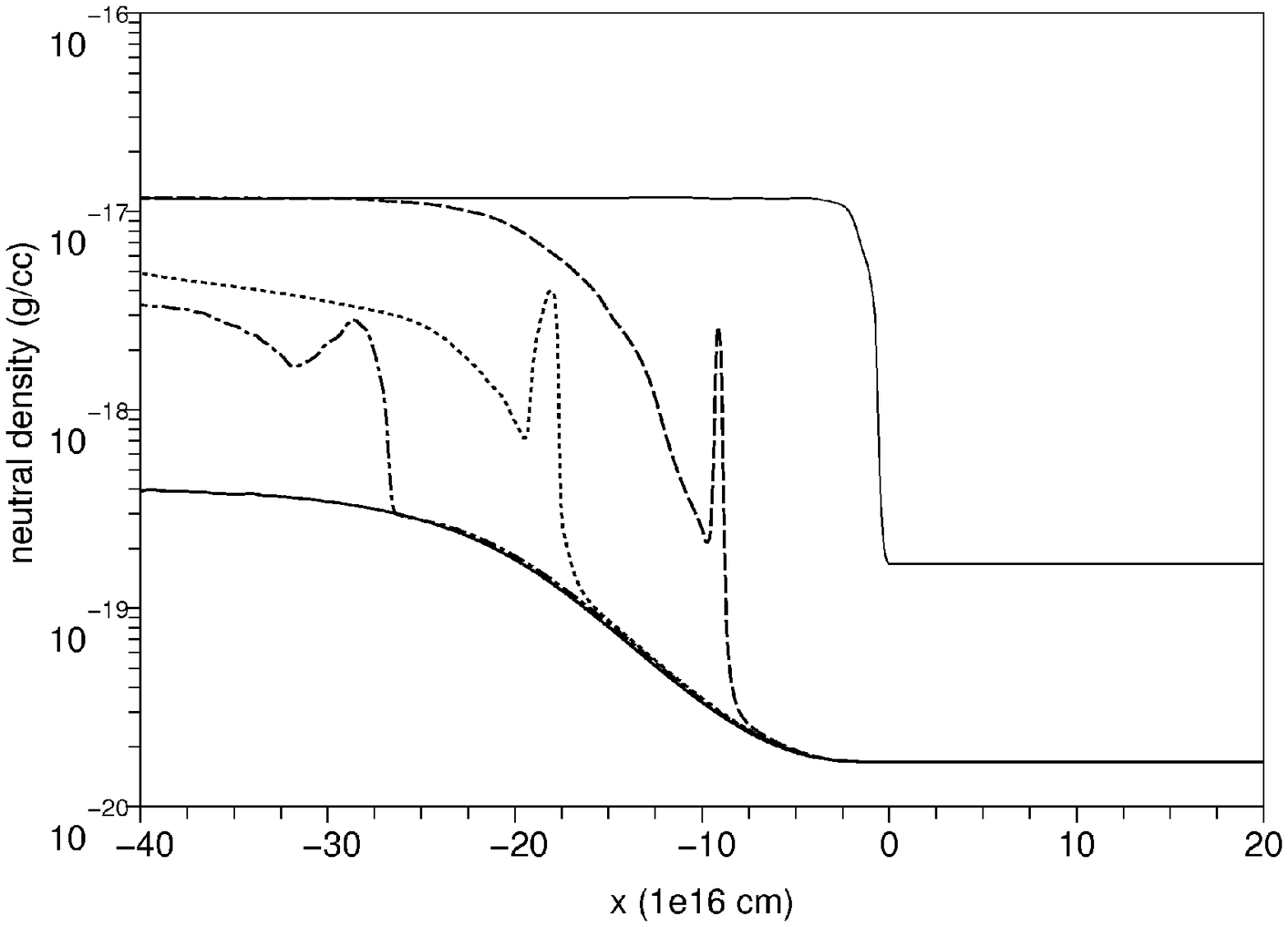}
\includegraphics[width = 8.4 cm]{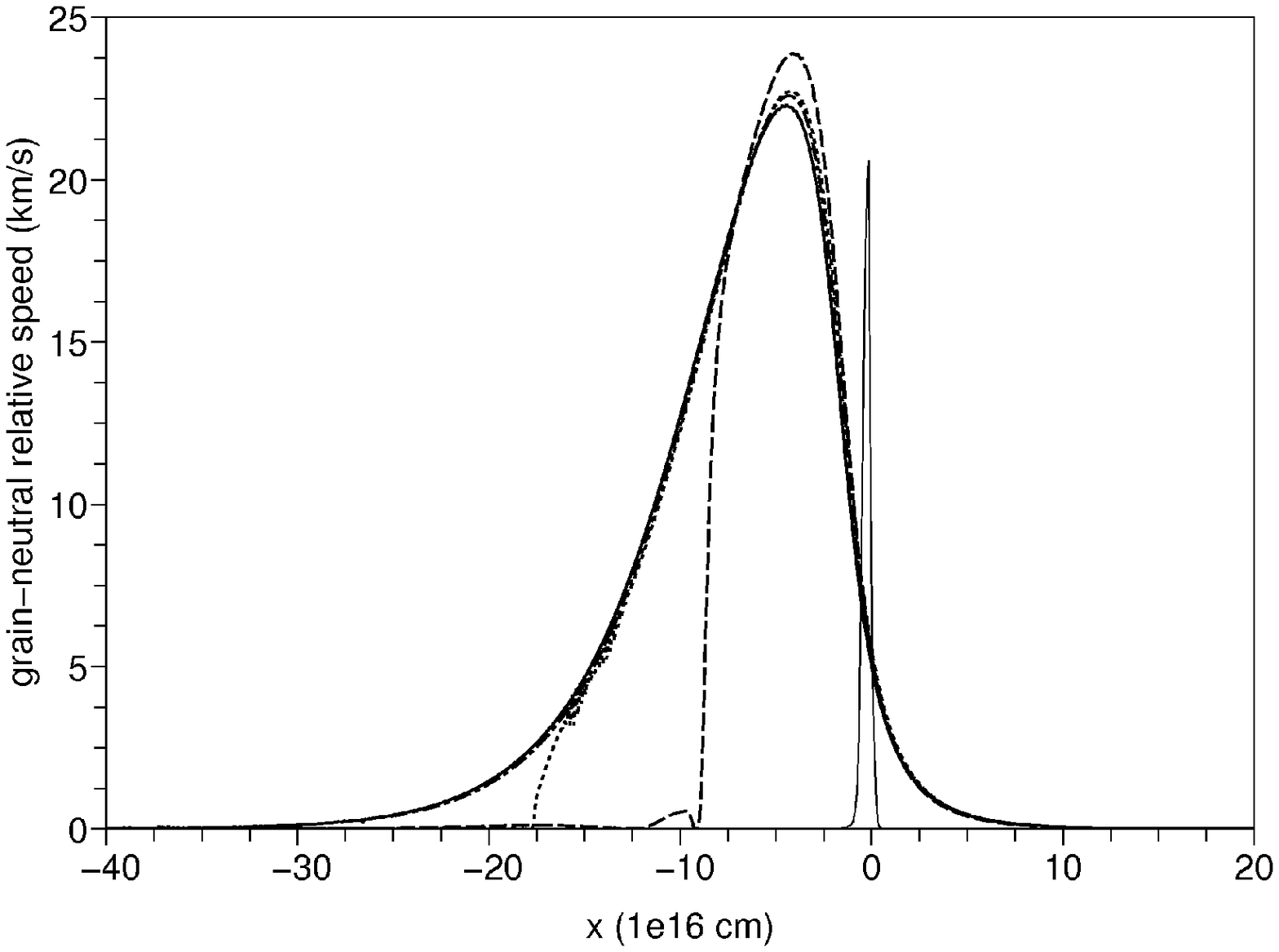}
\caption{Similar to Fig.~\ref{fig1}, but for a shock interacting with 
a decreasing density perturbation (from $n_{\rm H} = 10^5 {\rm cm^{-3}}$ to 
$n_{\rm H} = 10^4 {\rm cm^{-3}}$). The thin solid line is the initial 
C-type shock, the dashed line the shock structure $5\times 10^3$ yr after 
the shock interacts with  the density perturbation, the dotted  
line after $2.40 \times 10^{4}$ yr, the dash-dotted line  after 
$5.59 \times 10^{4}$ yr and the thick solid line the final steady C-type
shock (after $9.8 \times 10^{4}$ yr).}
\label{fig4}
\end{center}
\end{figure*}
 
In these early stages of the evolution we can already see the effect of the 
higher density on the charged particles fluid and the magnetic field. In the 
high density region the Hall conductivity becomes larger than the Pedersen 
conductivity (see definition in Paper I). This changes the wave propagation
upstream. This is associated with a change of the structure of the evolutionary
trajectory in the $B_y-B_z$ phase space \citep[cf.][]{W98}. When the Hall
conductivity exceeds the Pedersen conductivity, a spiral node develops in the 
trajectory for those regions corresponding to the flow around and just 
upstream of the leading part of the shock precursor.
The shock structure then contains large oscillations 
\citep[see][and Paper I]{F03} as seen in the $x$-component of the ion 
and electron velocities in Fig.~\ref{fig2}.  

The J-type phase of the shocks is quite short. Within $7.92 \times 10^4$ yr 
after impact, corresponding to 2 ion shock crossing times, the 
transmitted shock structure has 
become a stable C-type shock. For the reflected shock the time for the shock 
to become a stable C-type shock is somewhat 
shorter. The transmitted C-type shock is much narrower than 
the initial shock (roughly 1/8 of the original shock width). It also has 
a lower propagation speed of 16.3~km~s$^{-1}$ as the shock slows down when  
moving into a region of higher density and pressure. 

The density and velocity plots in Fig.~\ref{fig1} seem to suggest that the 
neutral flow in sufficiently upstream parts of the 
precursor is quasi-steady. The precursor shows some small oscillations in both 
width and gradient, but these are associated with numerical uncertainties. 
The quasi-steady evolution does not start immediately after the 
shock interacts with the density perturbation, but only commences after 
$7 \times 10^3$~yr. We note that this is roughly $(L_s + L_p)/V_s$ with 
$L_s$ the initial shock thickness and $L_p$ the perturbation width. 
It is the time needed for the shock and perturbation to cross 
each other.

Shocks with different propagation speeds or moving into regions with 
different density contrasts have behaviours similar to that of the model 
shock described above. Thus, such simulations give us some idea about
the relative strengths of the reflected-transmitted shock pairs and 
the timescales of the J to C-type transitions for cores in which the 
shock moves into a region of enhanced density. 

For instance, for a lower density contrast across the perturbation, the shock 
does not slow down as much as before. For example, the propagation speed of 
a shock with an initial speed of 25 km\ s$^{-1}$ is 
21.1 km\ s$^{-1}$ in a region with a density contrast of 3.
The transmitted shock is only marginally weaker than the 
initial C-type shock with peak temperatures of a few times 10$^4$~K. 
The timescale for the transmitted shock to become a steady C-type shock 
is slightly shorter than before (6.02 $\times 10^4$ yr or 
about 1.15 ion shock crossing times).
The post-shock density behind the transmitted shock is not much higher than 
the far downstream density leading to a weak reflected shock. The 
peak temperature in the reflected shock is only about 300~K, while 
it is of the order of a few times 10$^3$~K for a density contrast of 10. 
As this temperature is below the threshold for electron cooling to be 
important, the ion and electron temperatures in the reflected shock are 
similar. Furthermore, the grain-neutral relative velocity, which roughly 
scales with the temperature, in the reflected shock is too low for grain 
sputtering to be important, i.e. $|v_g - v_n| < 2\ {\rm km\ s^{-1}}$
(see discussion in Sect.~\ref{sect:conc}).
 
We also performed a simulation for an initial shock speed of 6 km\ s$^{-1}$
and a density contrast of 10. In this simulation the shocks were relatively 
weak.
While strong fast-mode shocks considerably heat the incoming gas, weak 
fast-mode shocks do not. For a shock velocity of 6 km\ s$^{-1}$ (or 
$M_A \approx 3$), the gas temperatures are below 10$^3$~K. The peak 
temperature in the transmitted and reflected shock are even 
less and are about 400 and $80$~K respectively. Note that the 
ratio of the temperatures is similar to the one for the 25 km\ s$^{-1}$ 
shock. Because of the low temperatures in the shock, the grains do not
get highly charged and their Hall parameter is of the order of unity.
The grains, therefore, move at a speed different from those of the neutrals and 
of the ions and electrons. As the relative grain-neutral velocity is 
quite small ($\ll$ 1 km s$^{-1}$), grain sputtering will not occur. 
Variations in the transmitted-shock structure dissappear after 
9.20 $\times 10^4$ yr (about one ion shock crossing times) when the shock 
reaches steady state. An interesting difference between this model and 
the strong fast-mode shock 
models is that the transmitted shock width is not much smaller than of the 
width of the initial C-type shock. The ratio of the shock widths is 0.62. 
This is not 
surprising as the fractional ionisation does not change much across the 
density perturbation due to the low temperatures of the shock. Another 
difference is that, for the weak shock case, the trajectory in the $B_y-B_z$
phase space has no spiral node.  This 
is because,  as the grains are not important charge carriers, 
the Hall conductivity does not modify the upstream conditions.

\subsection{Decreasing density}\label{subsec:consthigh-low}

\begin{figure*}
\begin{center}
\includegraphics[width = 7.6 cm]{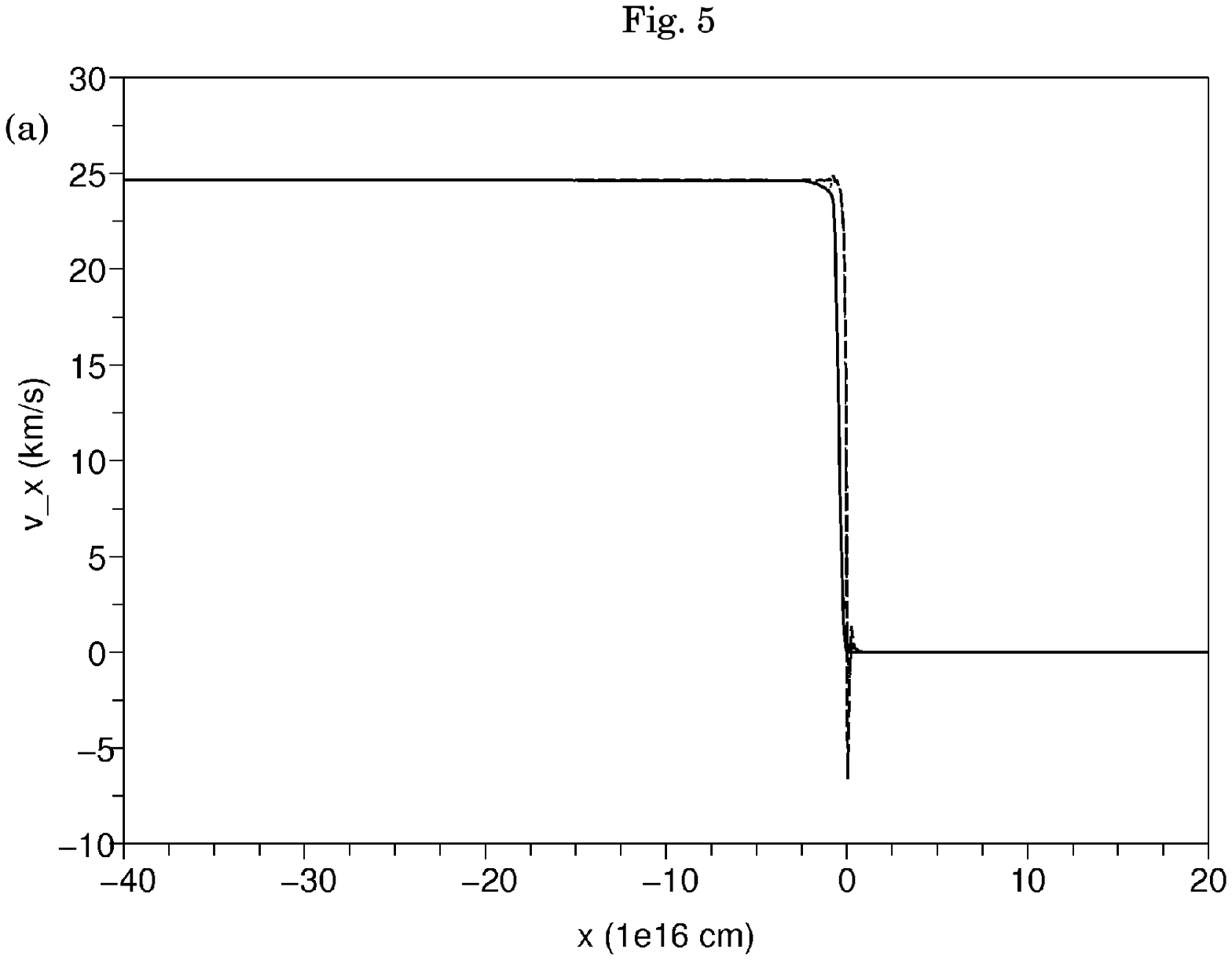}
\includegraphics[width = 7.6 cm]{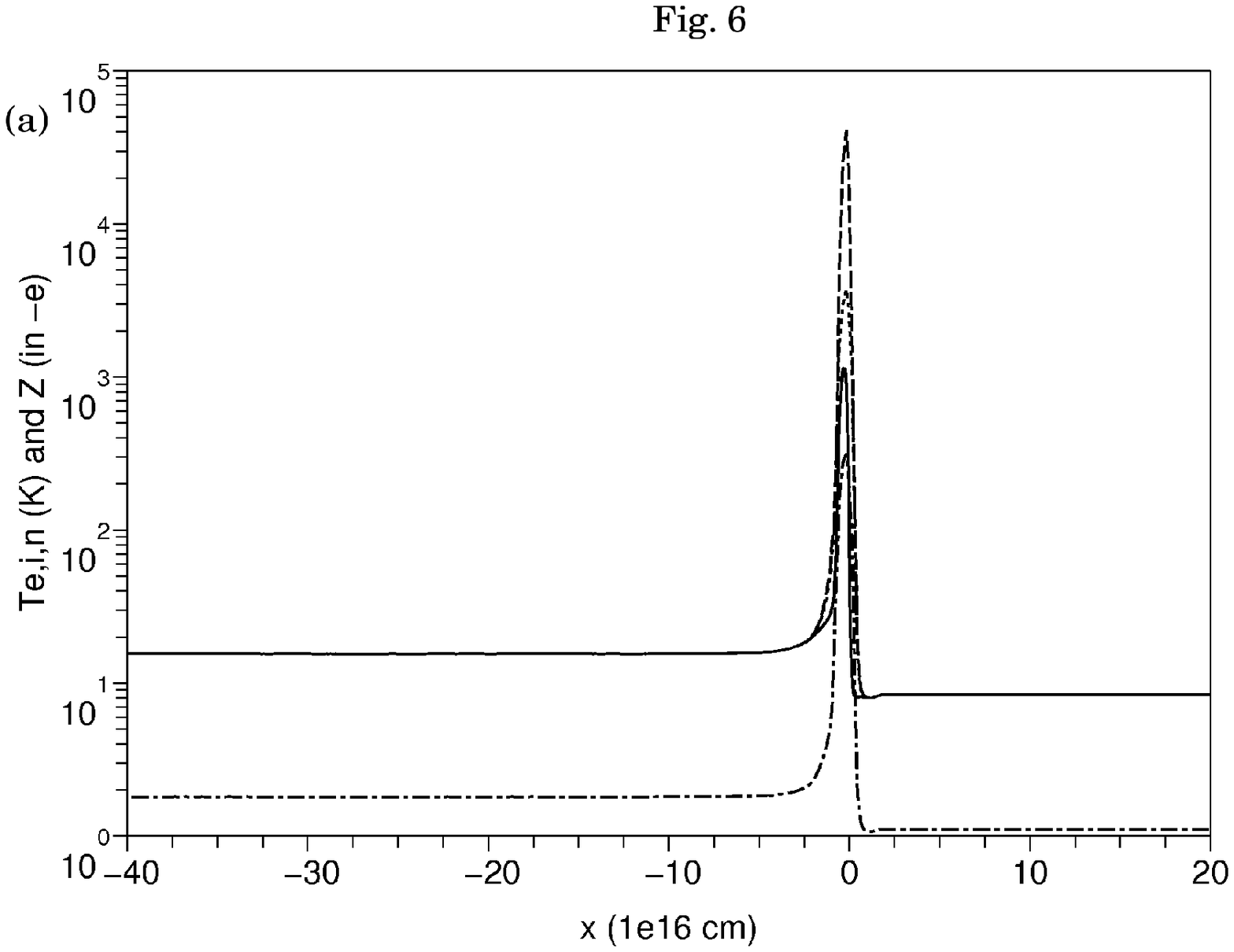}
\includegraphics[width = 7.6 cm]{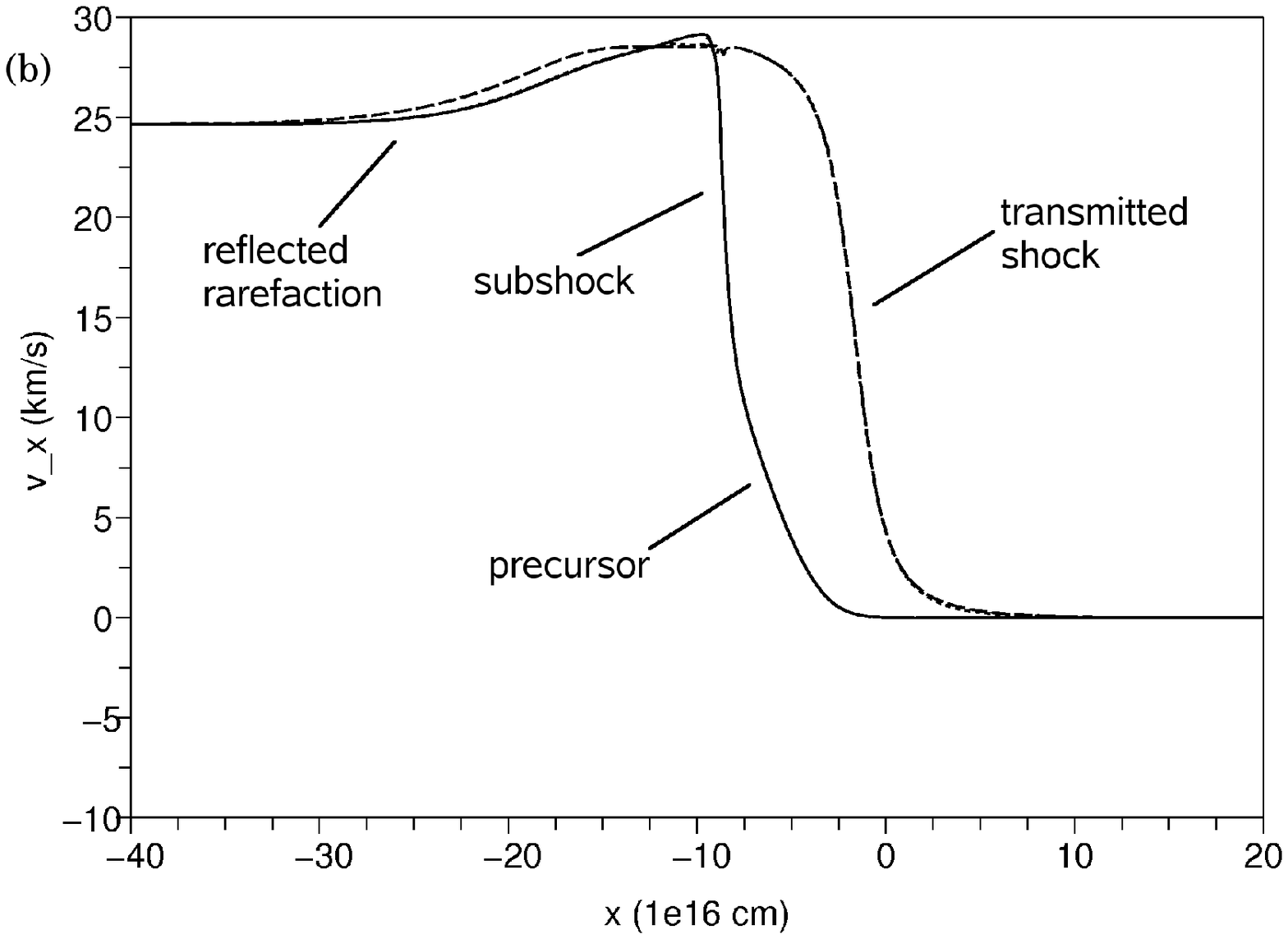}
\includegraphics[width = 7.6 cm]{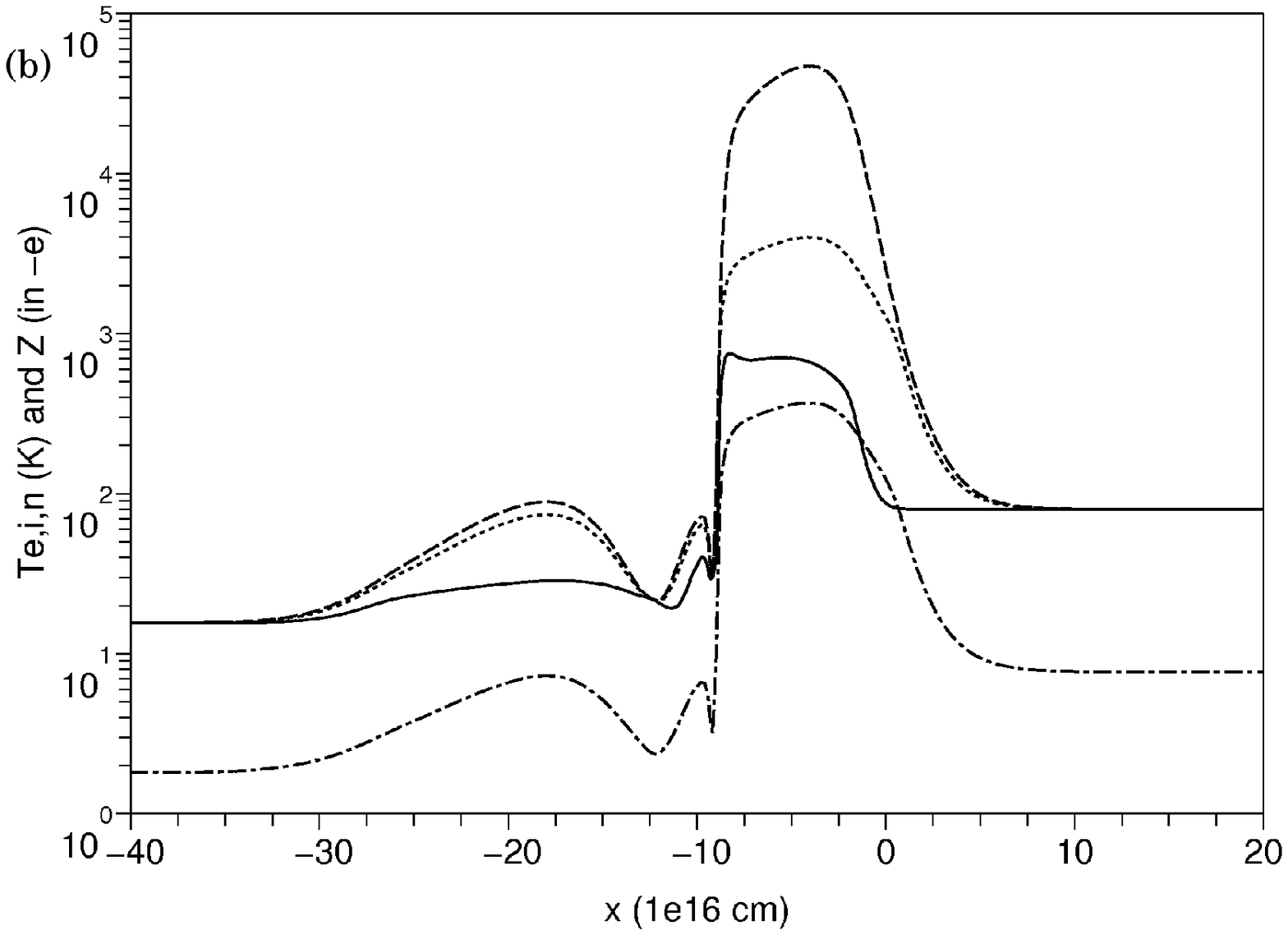}
\includegraphics[width = 7.6 cm]{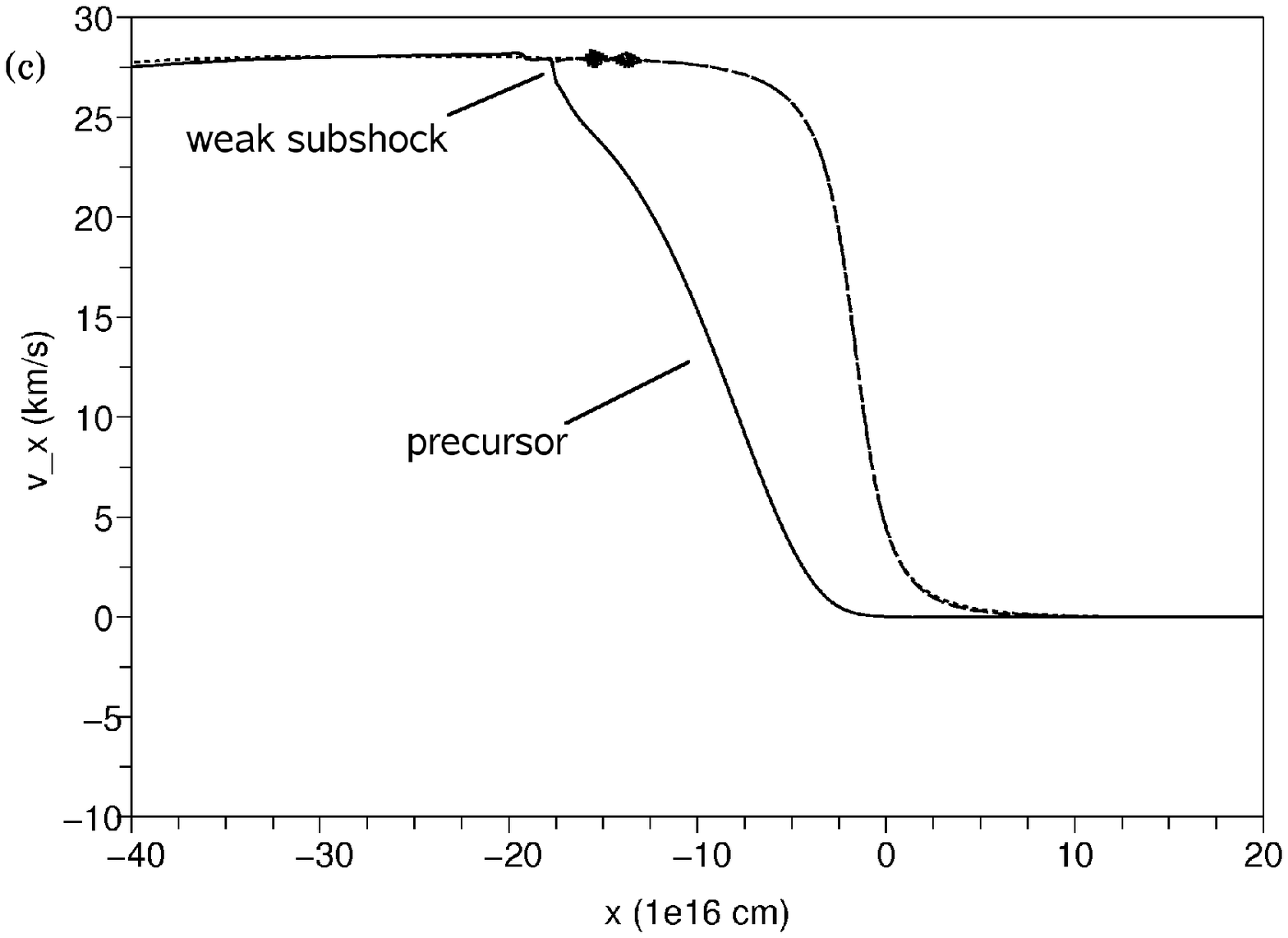}
\includegraphics[width = 7.6 cm]{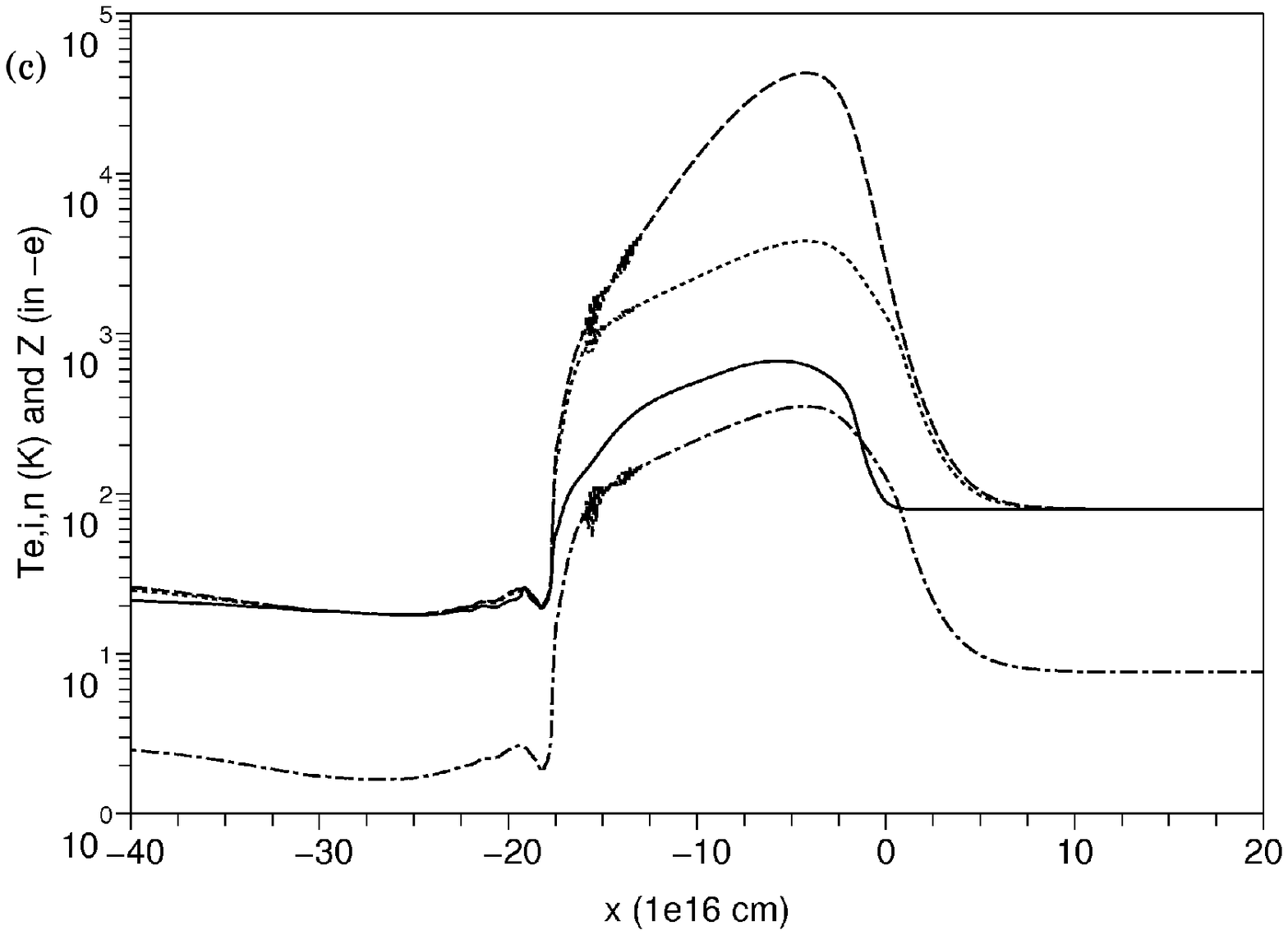}
\includegraphics[width = 7.6 cm]{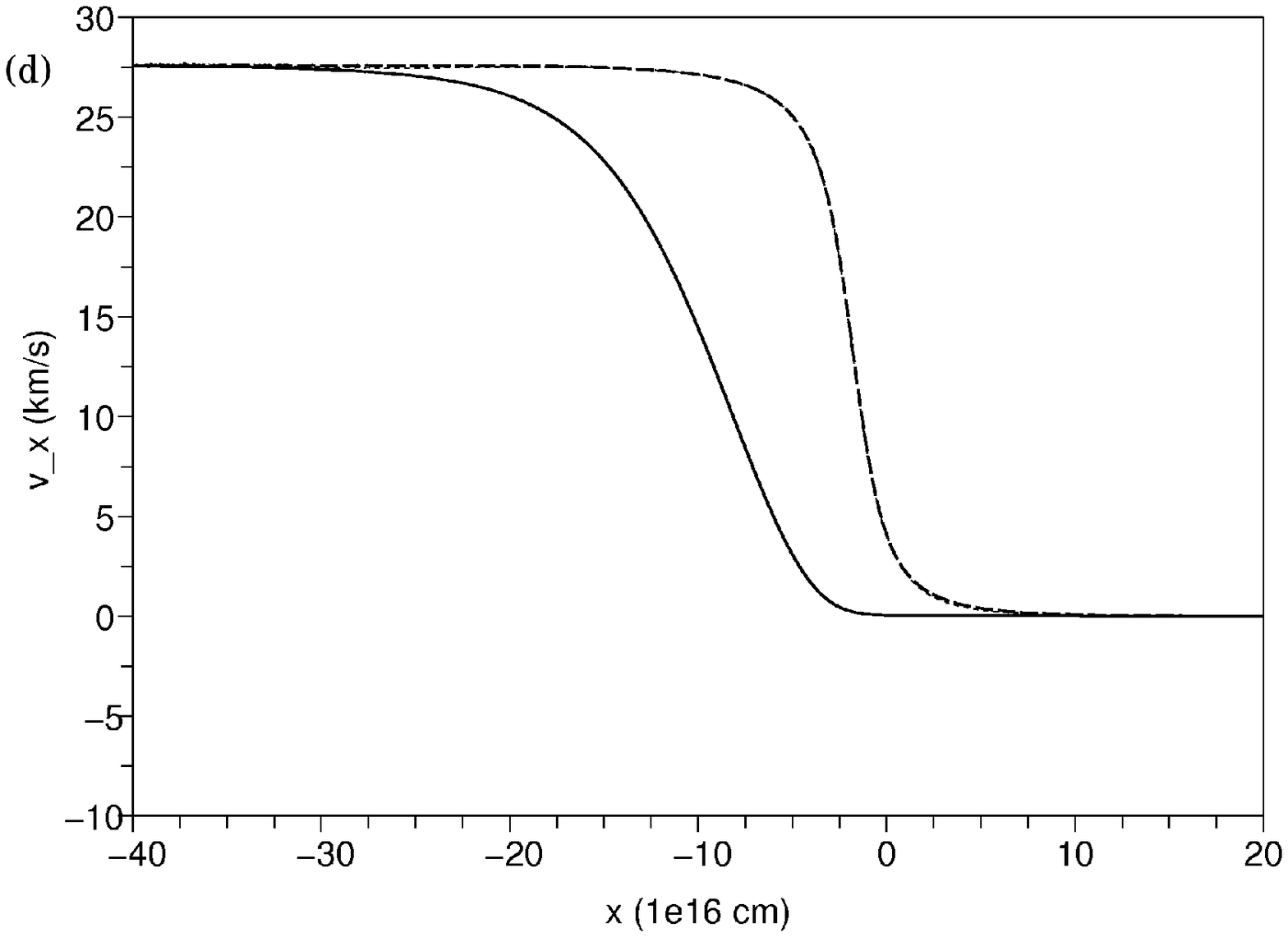}
\includegraphics[width = 7.6 cm]{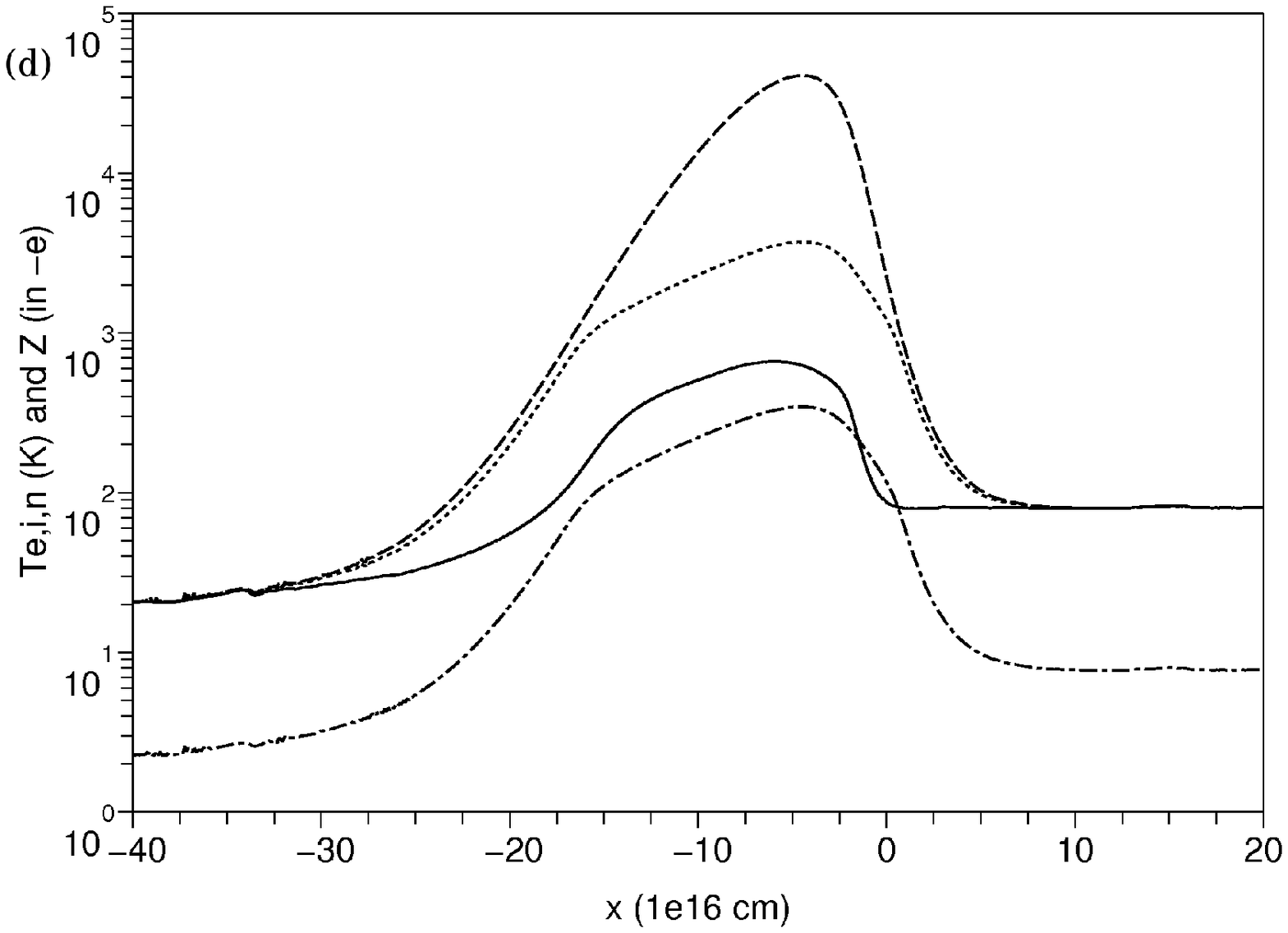}
\caption{Similar to Fig.~\ref{fig2}, but for a shock interacting with 
a decreasing density perturbation (from $n_{\rm H} = 10^5 {\rm cm^{-3}}$ to
$n_{\rm H} = 10^4 {\rm cm^{-3}}$). (a) shows the initial condition, 
(b) the shock structure $5 \times 10^3$ yr after the interaction
with the density perturbation, (c) after  
$2.40 \times 10^{4}$ yr and (d) the final steady state shock (after
$9.8 \times 10^4$ yr).}
\label{fig5}
\caption{Similar to Fig.~\ref{fig3}, but for the shock and times shown 
in Fig.~\ref{fig5}.}
\label{fig6}
\end{center}
\end{figure*}

For this model, a 25 km\ s$^{-1}$ C-type shock propagates out of 
a dense region with $n_{H} = 10^5~{\rm cm^{-3}}$ into a more diffuse one with
$n_{H} = 10^4~{\rm cm^{-3}}$. Such a model is representative of 
a molecular outflow leaving the dense core surrounding a protostellar 
object. 

Note that the final steady shock for the case to which Fig.~\ref{fig1} 
corresponds, is similar to the initial shock structure here (see 
Figs.~\ref{fig4}, \ref{fig5} and \ref{fig6}).  Many features of the shock-clump
interaction are reversed. As the shock moves into the lower density region,
the shock speeds up and broadens (see Fig.~\ref{fig4}). Furthermore, 
the spiral $B_y-B_z$ feature does not appear. A rarefaction wave 
propagates into the 
far downstream gas. In this case, the rarefaction is a slow-mode wave
as the magnetic field decreases across the structure. As for multifluid
shocks, a multifluid rarefaction wave also introduces an ion-neutral drift 
(see e.g. Fig.~\ref{fig5}b). However, the relative velocities between the 
charged particles and neutrals is negligible compared to the velocity of 
the transmitted shock. 
 
In the transmitted shock a neutral subshock and a precursor are 
initially present. From the precursor, a C-type shock eventually develops. 
As in the increasing density, the quasi-steady evolution starts after 
an initial adaptation phase of the order 
1.7$\times 10^3$~yr ($=(L_s + L_p)/V_s$).
The transmitted shock reaches steady state after 
$9.80 \times 10^4$ yr, which roughly corresponds to 
1.2 times the ion flow time through the final shock structure. 

As mentioned before, the transmitted shock speeds up to 
$\approx 28.6$ km\ s$^{-1}$ or $M_A \approx 13$. The shock is thus a
strong shock and the gas temperatures are high within the shock. (I.e. 
the ions have peak temperatures of 10$^5$~K.) As the high temperatures are 
due to collisions between the different fluids, the ion-neutral and 
grain-neutral relative velocities within the shock structures also remain 
high (see Fig.~\ref{fig4}). Furthermore, the spatial region with 
high relative velocities is larger than for the case for which results are 
shown in Fig.~\ref{fig1}, which suggests that grain sputtering is 
going to be more effective than in the initial C-type shock.

\subsection{Sinusoidal density perturbation}\label{subsec:sinclump}

\begin{figure*}
\begin{center}
\includegraphics[width = 8.4 cm]{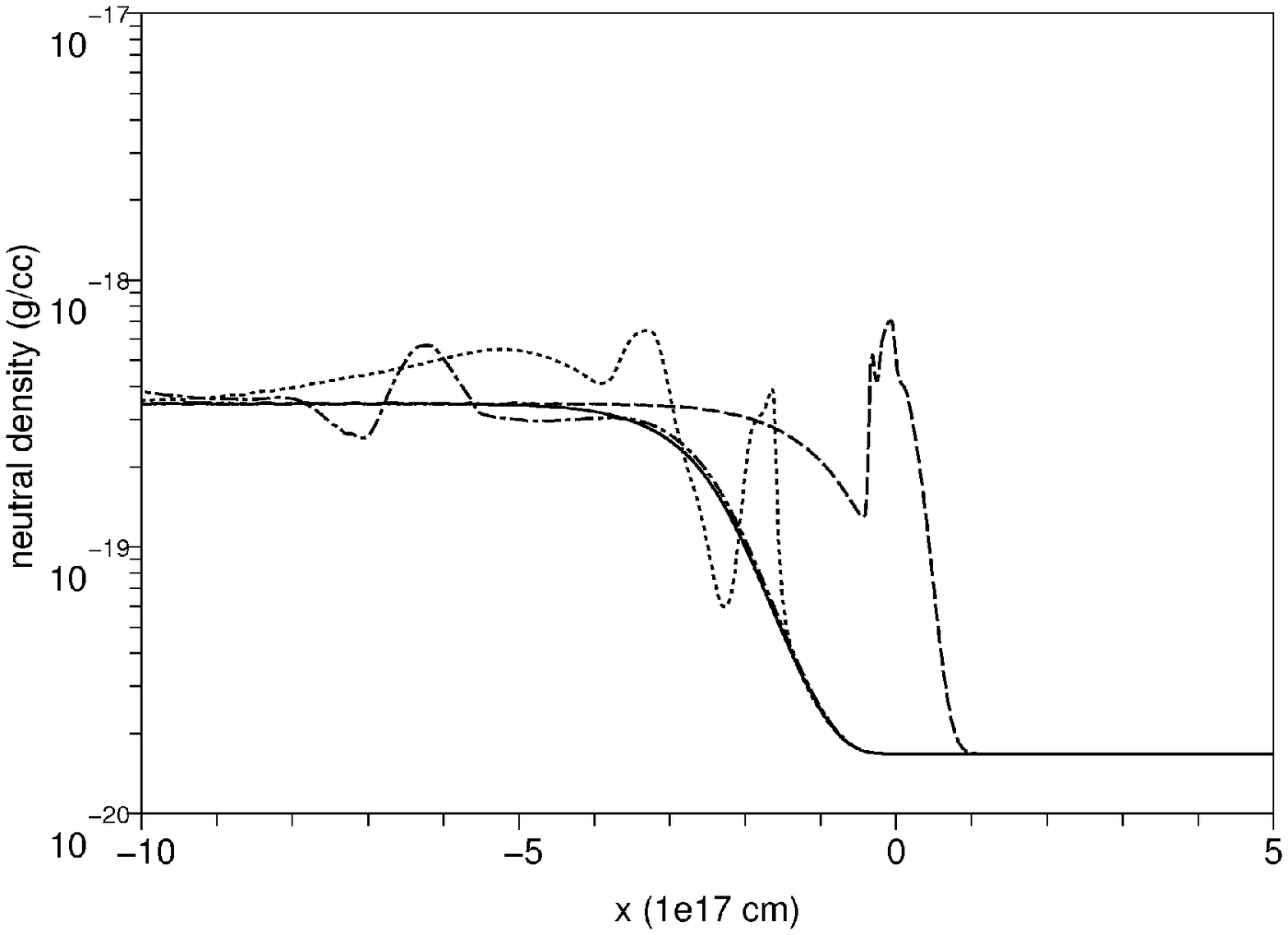}
\includegraphics[width = 8.4 cm]{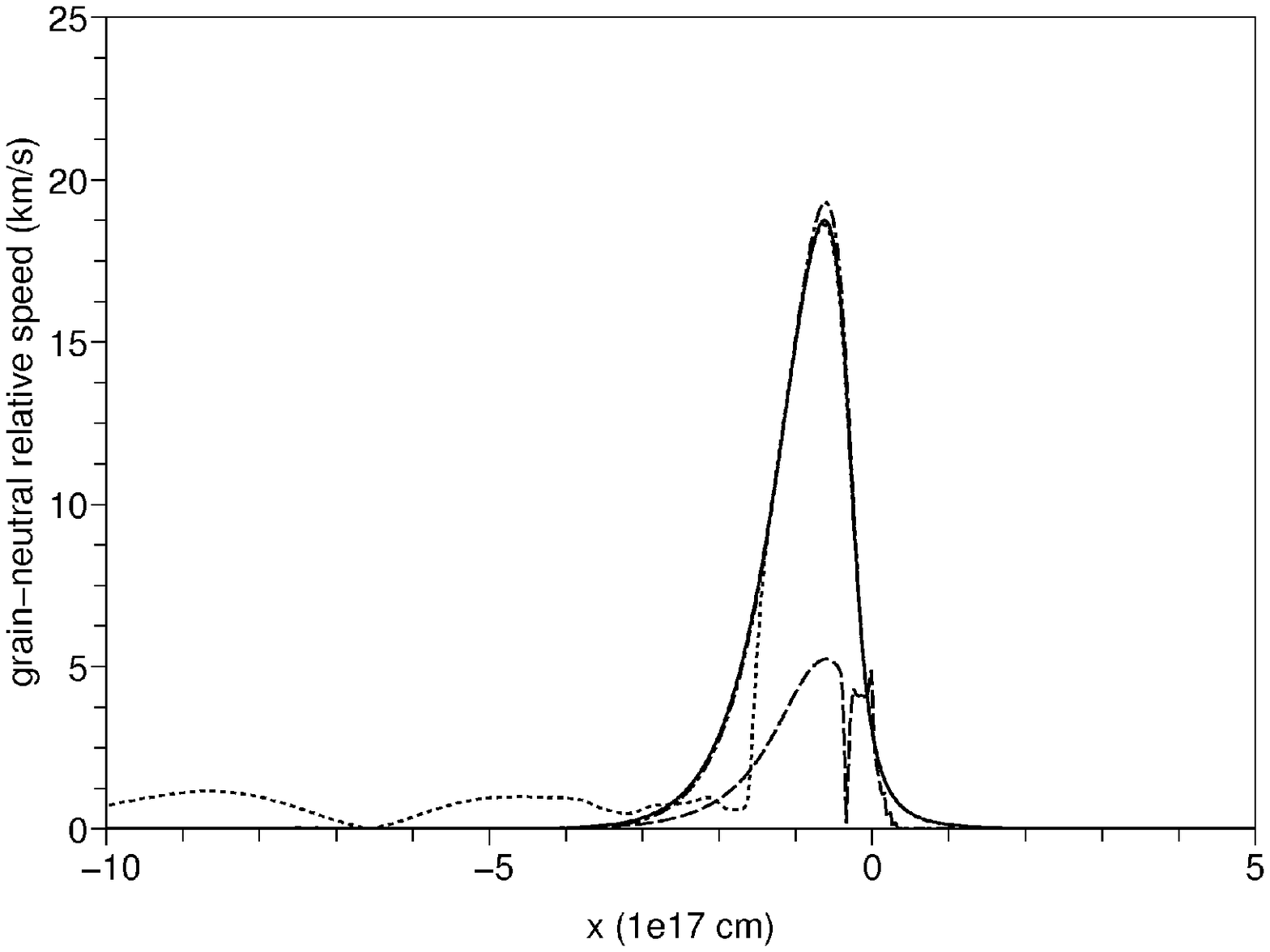}
\caption{Similar to Fig.~\ref{fig1}, but for a shock interacting with 
a clump that has a maximum density of $n_{\rm H} = 2 \times 10^5 {\rm cm^{-3}}$.
The solid line shows the initial and final steady shock (after 
1.2 $\times 10^5$~yr), the dashed line the shock structure 
3.9 $\times 10^3$~yr after the shock interacts with 
the clump, the dotted line after 1.98 $\times 10^4$~yr and the dash-dotted 
line after 1.46 $\times 10^5$~yr.}
\label{fig7}
\end{center}
\end{figure*}

\begin{figure*}
\begin{center}
\includegraphics[width = 7.6 cm]{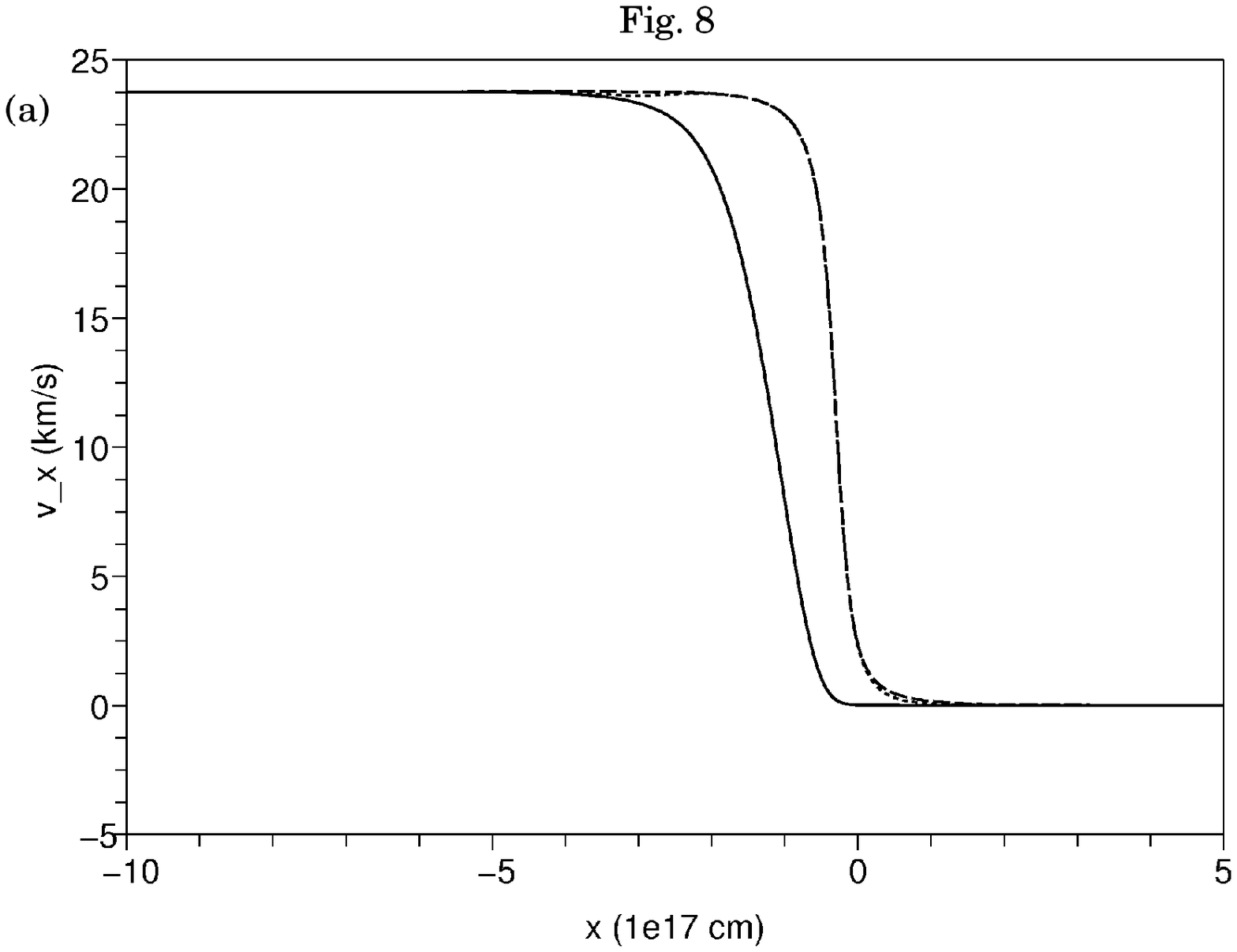}
\includegraphics[width = 7.6 cm]{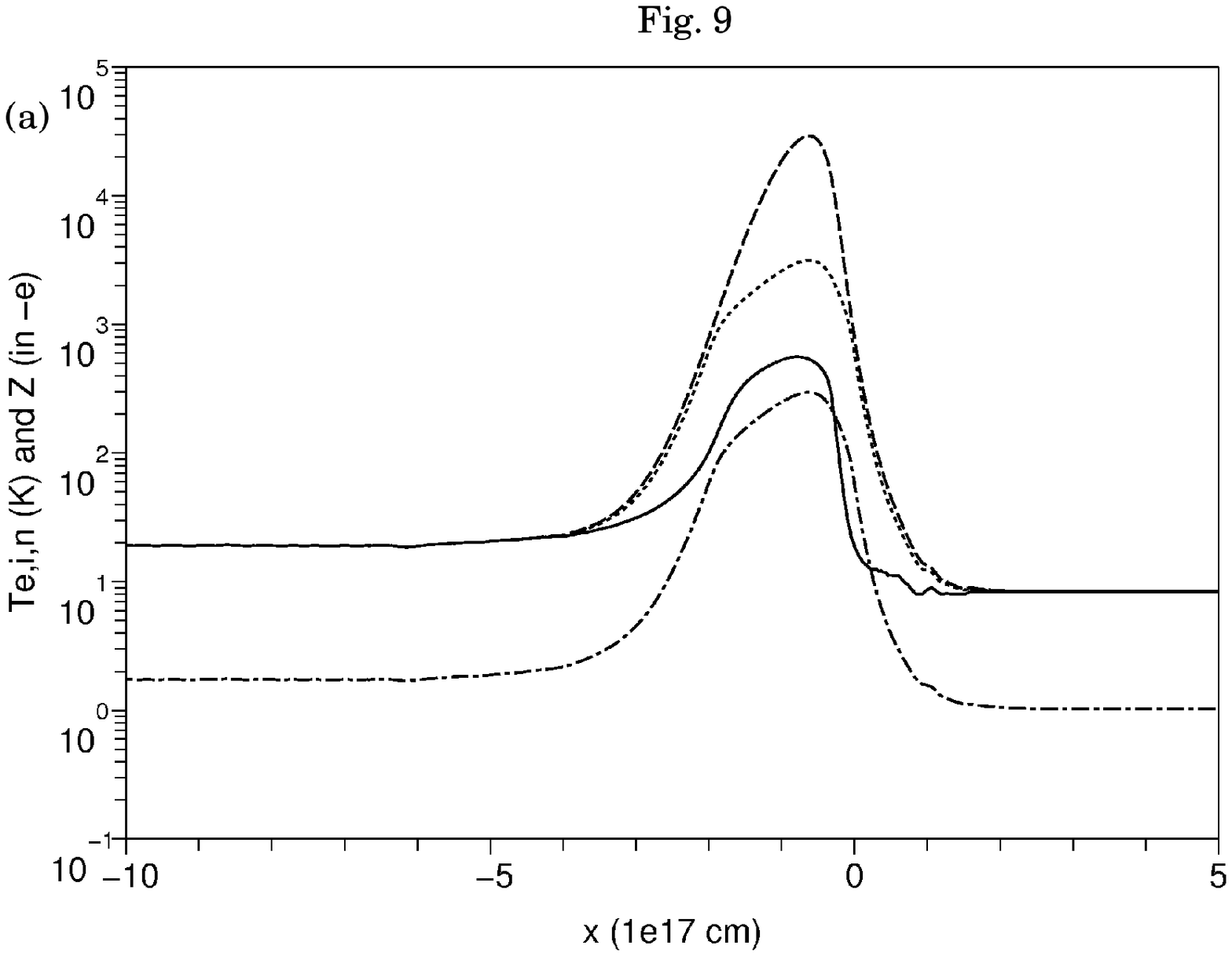}
\includegraphics[width = 7.6 cm]{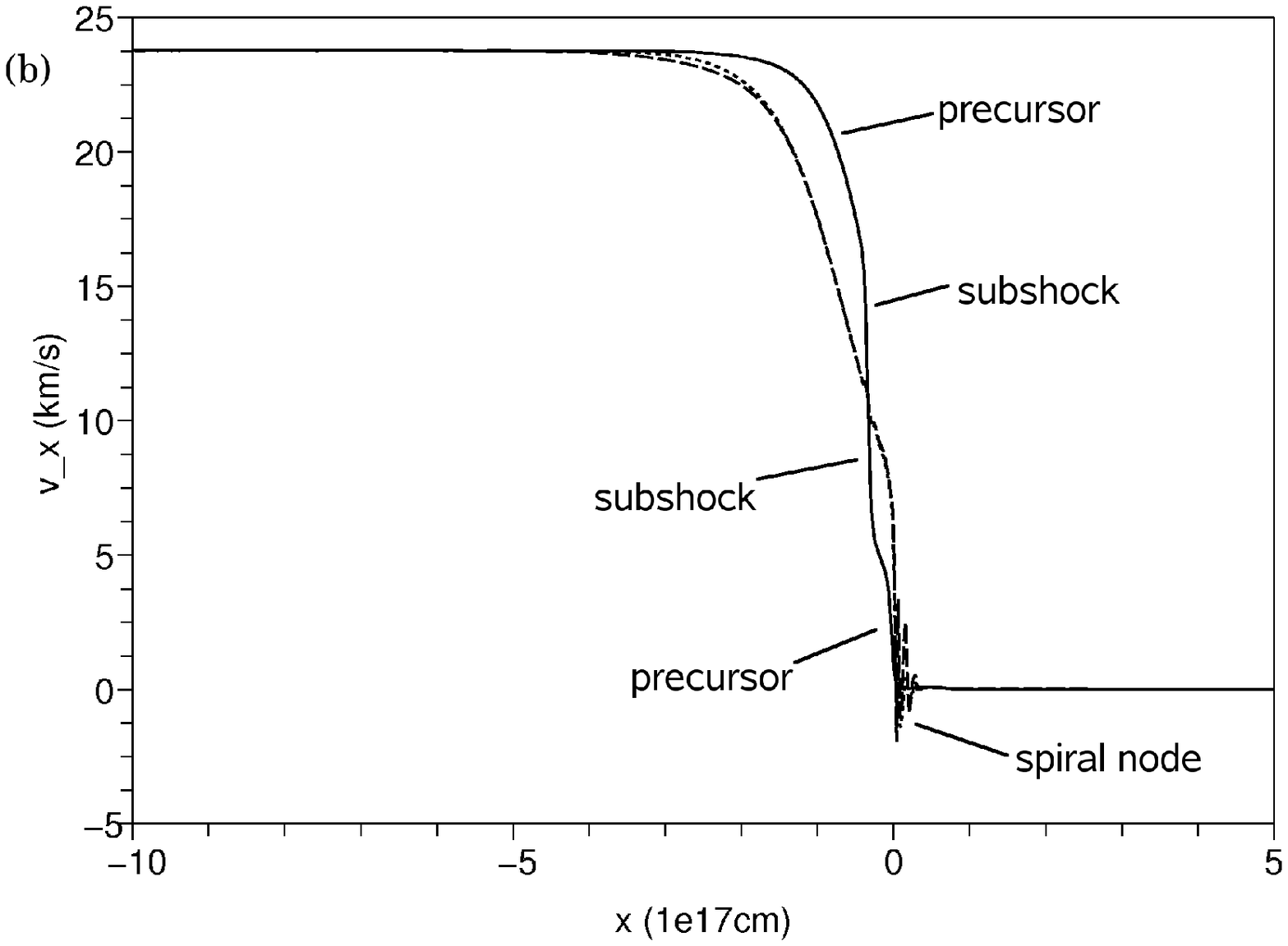}
\includegraphics[width = 7.6 cm]{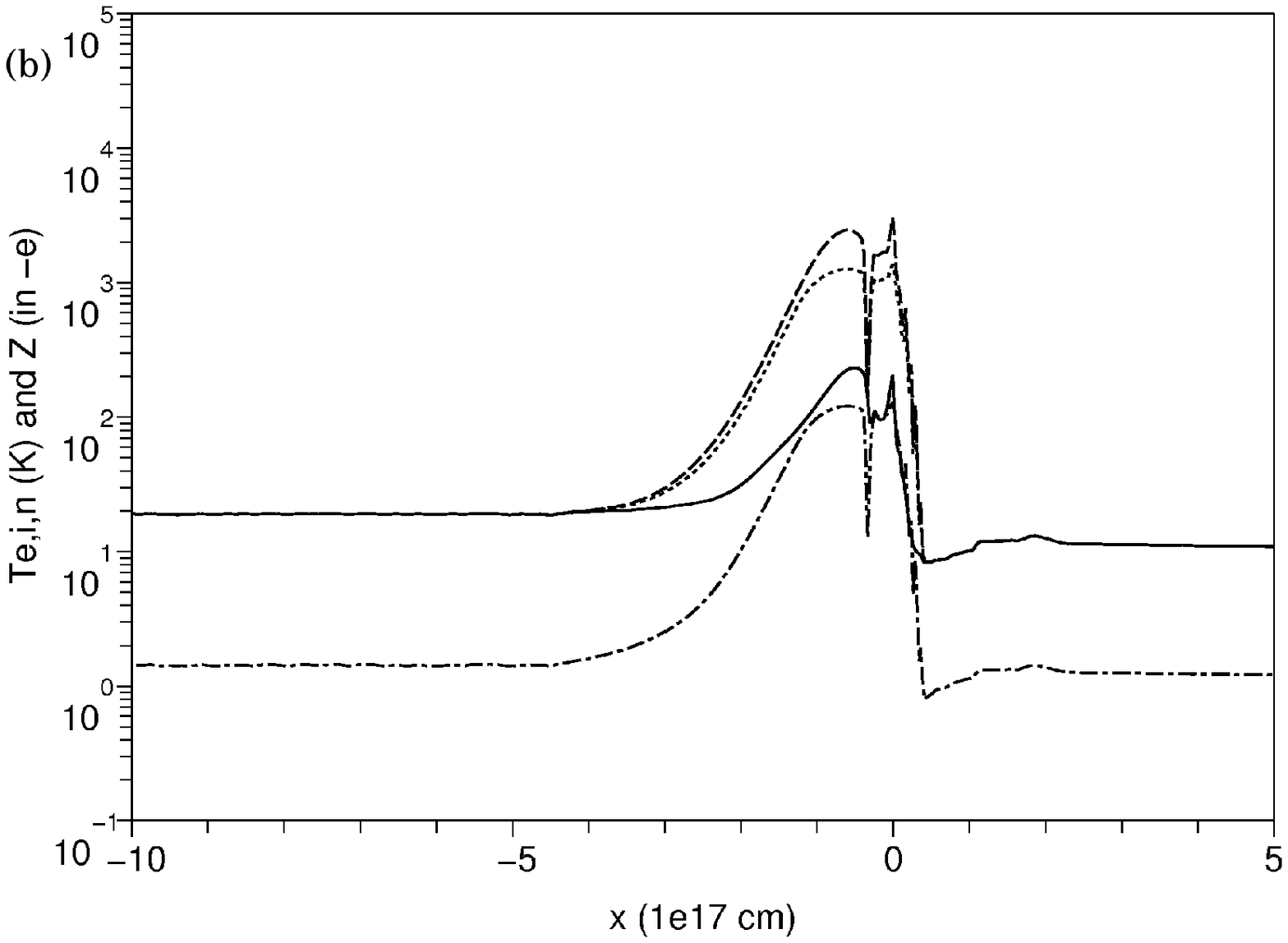}
\includegraphics[width = 7.6 cm]{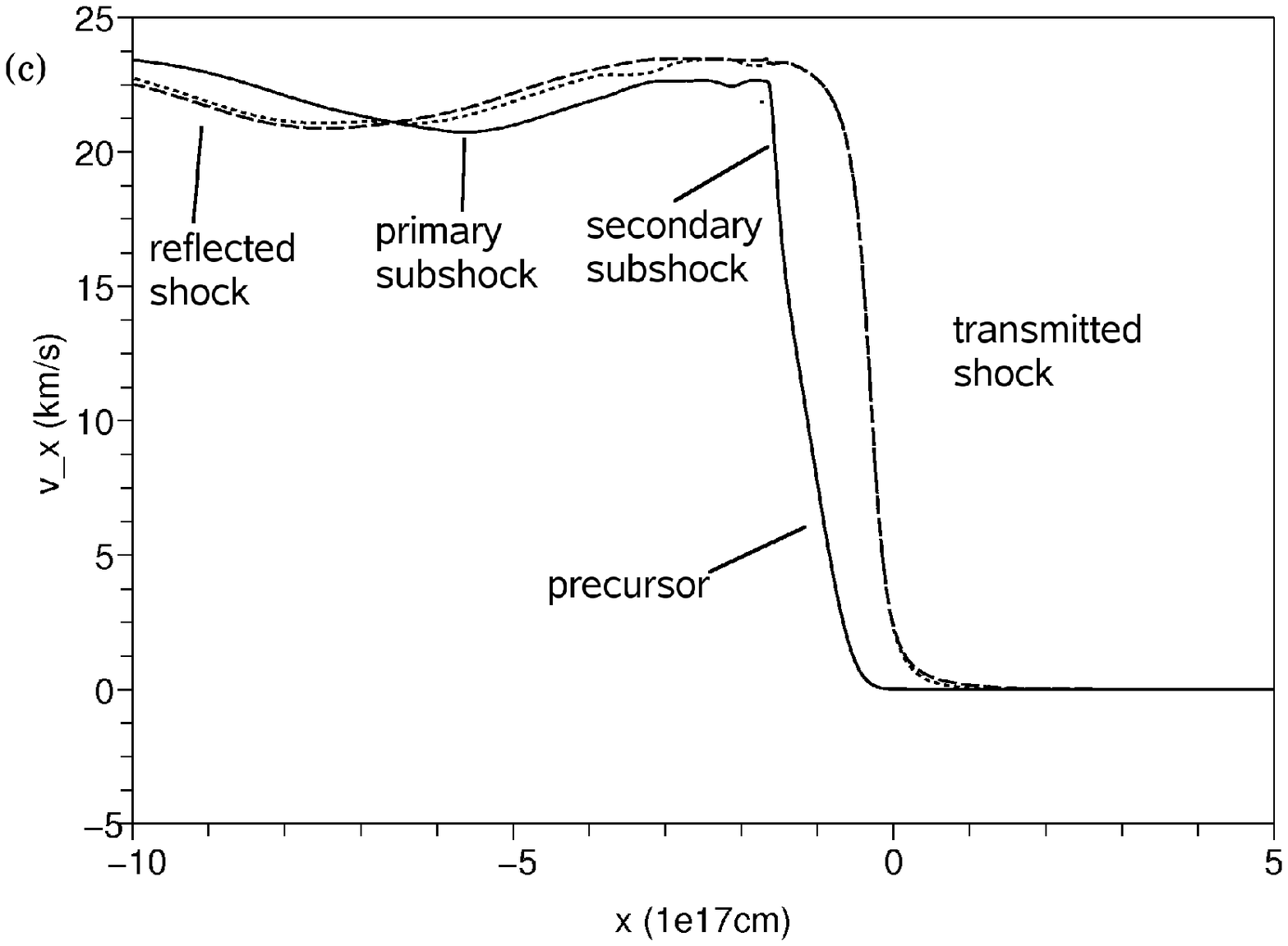}
\includegraphics[width = 7.6 cm]{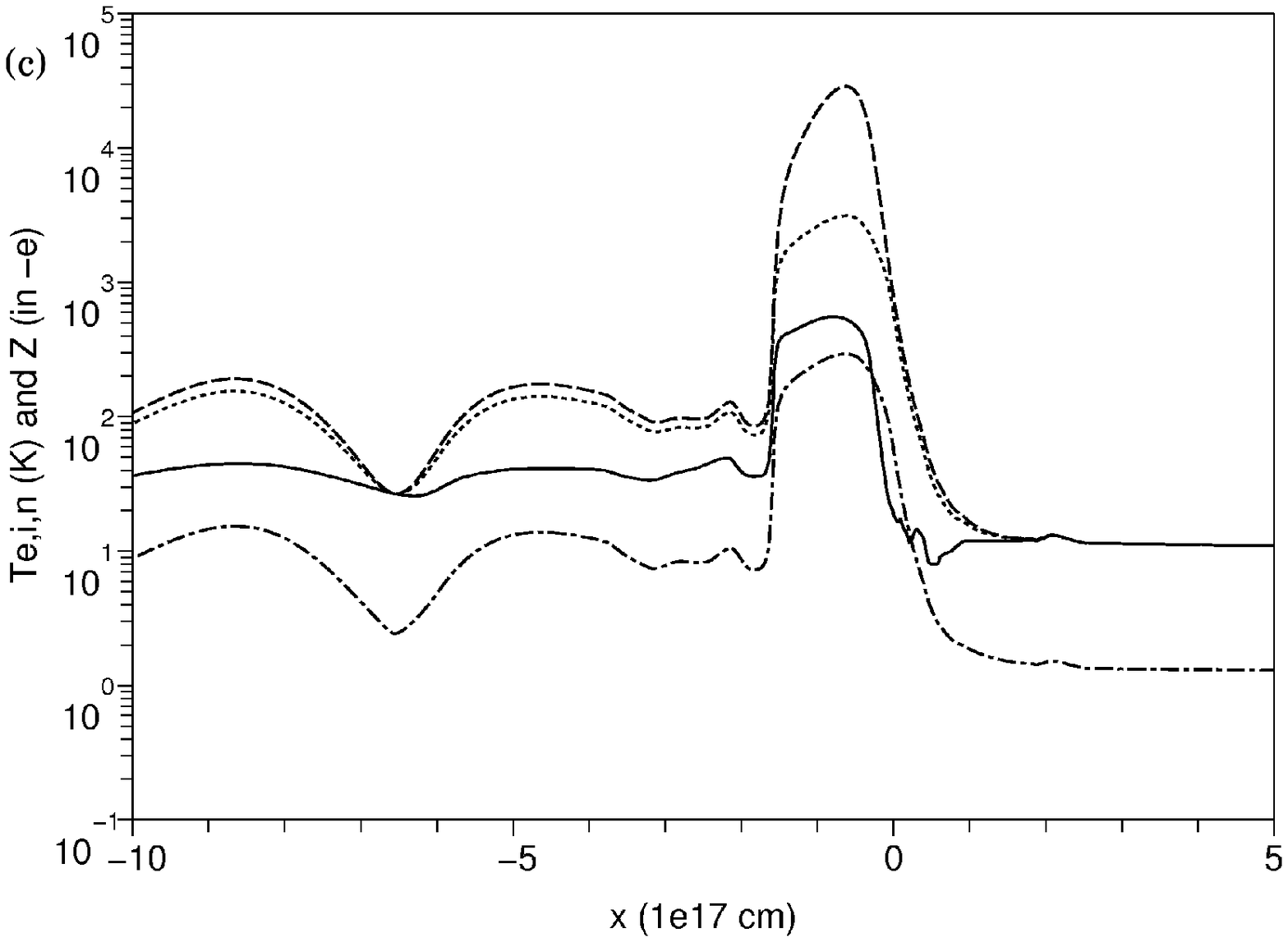}
\includegraphics[width = 7.6 cm]{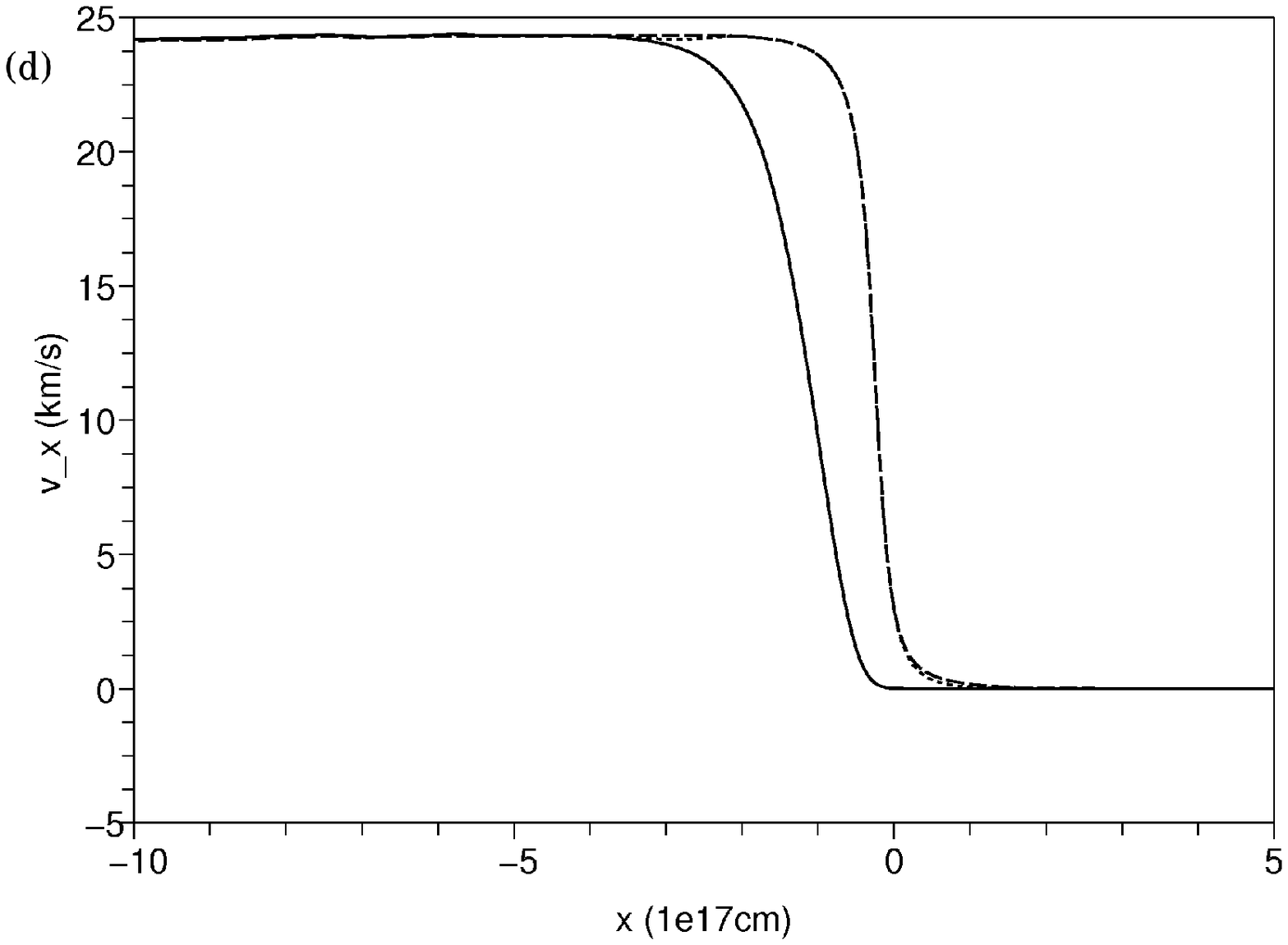}
\includegraphics[width = 7.6 cm]{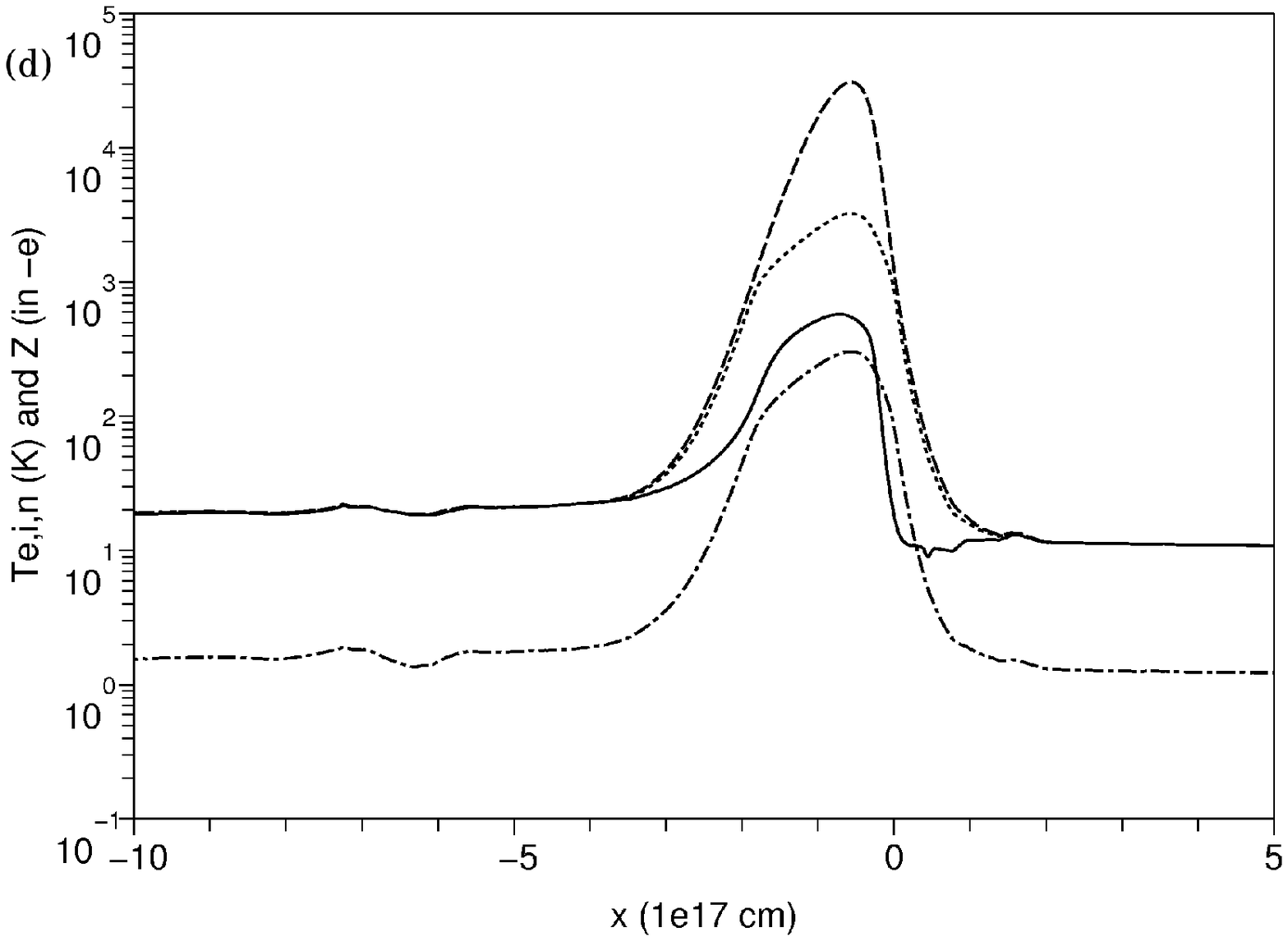}
\caption{Similar to Fig.~\ref{fig2}, but for a shock interacting with
a clump that has a maximum density of $n_{\rm H} = 2 \times 10^5 {\rm cm^{-3}}$.
(a) shows the initial and final shock structure (after 1.2 $\times
10^5$~yr), (b) the shock structure 3.9 $\times 10^3$~yr after the shock first 
interacted with the clump, (c) after 1.98 $\times 10^4$~yr 
and (d) after 1.46 $\times 10^5$~yr.}
\label{fig8}
\caption{Similar to Fig.~\ref{fig2}, but for the shock and times shown in 
Fig.~\ref{fig8}.}
\label{fig9}
\end{center}
\end{figure*}

We also study the interaction of a C-type shock with a clump. 
The shape of the clump is given by a cosine wave with a wavelength 
($\lambda$) of 1.46 $\times 10^{17}$~cm, or more specifically 
\[
 	n_H = \left(\frac{n_{H,max} - n_{H,0}}{2}\right) 
	\cos \left[\frac{2\pi(x - x_0)}{\lambda} - \pi \right] + 
	\left(\frac{n_{H,max} + n_{H,0}}{2}\right),  
\]
with $x_0$ the start position of the clump. The maximum density in the clump is 
$n_{H,max} = 2 \times 10^5~{\rm cm^{-3}}$ and is 20 times higher than 
the background density ($n_{H,0}$). These values are representative for clump 
parameters measured by \cite{Tal04}.
 
Unfortunately, the evolution of the shock structure cannot be described 
by a combination of the previous two models. The behaviour 
of the shock as it moves up the density gradient is similar to the early
evolution shown in Fig.~\ref{fig1}.
The dashed line in Fig.~\ref{fig7} and in Fig.~\ref{fig8}b and~\ref{fig9}b
show that a reflected shock and a transmitted J-type shock form
when propagating up the density gradient. The transmitted shock slows 
and narrows. A spiral feature develops in the trajectory in $B_y-B_z$
phase space.

The motion back to the low density region does not resemble the transition of 
Sect.~\ref{subsec:consthigh-low}. This is because  
the transmitted shock does not become a steady state C-type shock 
within a few thousand years. (The time scale for the shock to adapt to 
the new upstream condition is by itself a few thousand years.) 
Rather, a J-type shock 
moves down the density gradient. As the  shock reacts to the 
changing upstream conditions, it broadens and speeds up. 
The precursor then develops a neutral subshock just as the C-type shock did 
for the case for which Fig.~\ref{fig4} shows results. This means that the 
shock structure simultaneously has two neutral subshocks (see 
Fig.~\ref{fig8}c). Interestingly,
the neutral and charged flows do not have the same post-shock
velocities behind the secondary subshock. However, this is not 
surprising as the relative 
velocity of the neutral and charged flow in the original precursor is  
non-zero.  

Although these relative velocities are well below the sputtering 
thresholds derived in \cite{Cal97}, other initial parameters may yield 
relative velocities above this threshold which extend over a spatial region
larger than the initial shock width. The relative velocities steadily
decrease as both subshocks eventually weaken. The initial one weakens after
a few 10$^4$~yr. Simultaneously, a steady C-type shock forms from the 
precursor (see Fig.~\ref{fig7}).  
The timescale for the steady C-type shock to arise is about 1.2 $\times 10^5$~yr
or 1.2 times the ion flow time through the shock. 
The final C-type shock is, as expected, identical to the 
initial one.

\section{Summary and discussions}\label{sect:conc}
In this paper we present the first time-dependent simulations of 
oblique C-type shocks in dusty plasmas interacting with density perturbations.
We studied the transient evolution of the shock structure for each of several
different 
types of density perturbations, i.e. increasing, decreasing and clump-like,
and examined the dependence of this interaction on shock speed and 
density contrast. 

When a steady C-type shock encounters a density perturbation, 
the shock reacts to the changing upstream condition by breaking up into
multiple waves. While a J-type shock moves into or out of a dense
region, a shock or rarefaction wave, respectively, is reflected and propagates
in the reverse direction. This reflected shock or wave 
is present because the post-shock flow behind the transmitted shock is different
from the post-shock flow of the initial shock. 
The relative strength of the transmitted and reflected components depends
on the initial shock speed and the density contrast of the perturbation, 
as these control  the post-shock properties of the transmitted shock, e.g. 
a lower density 
contrasts leads to a smaller difference between the far downstream flow and the 
transmitted post-shock flow. Hence, a weaker rarefaction wave or shock is needed
to adjust this change. For the interaction with a clump, where the density 
decreases again, the situation gets more complicated as the transmitted 
J-type shock forms a second subshock in the neutral flow of the precursor. 
It is thus important to realise that we expect a range of different 
J- and C-type shocks and multifluid rarefaction waves in inhomogeneous
clouds.  

In our models, we have focused on the evolution of the transmitted 
shocks. Such a shock is initially J-type and contains a subshock in the neutral
flow. It again becomes a steady 
C-type shock but with a different shock width and velocity. The 
timescale for it to become steady is of the same order as the ion-flow time
through the final shock structure, i.e. about 0.5-1.5 $\times 10^5$ yr.
An important consequence of these timescales is that, if a shock interacts 
with density inhomogeneities on timescales shorter than this, no steady
C-type shock exists. All shocks are thus J-type, although some shocks will 
have weak subshocks and are then only marginally distinguishable from C-type
shocks.

Our simulations show that the steady dusty C-type shock 
develops from a shock in which the sufficiently far upstream parts of the 
precursor are steady. Such an evolution was also observed for multifluid
shocks in dustless, weakly-ionised gases \citep{FP99}. Following 
the \cite{FP99} 
result, \cite{Lal04} developed a quasi-steady method to follow the temporal 
evolution of multifluid shocks. This method involves the treatment of an 
evolving non-stationary J-type shock as a sequence of truncated C-type shocks 
each of which has a neutral subshock at the point where the ion-flow time 
corresponds to the age of the shock. While this approach was only validated 
for shocks in dustless, weakly ionised gases \citep{Lal04}, our result 
confirms quasi-steady models can also be used for dusty shocks and justifies 
the use of such methods for the temporal evolution of dusty shocks 
by \cite{Getal08}. However, we also find some limitations of the 
quasi-steady approach when applied to the interaction of shocks with density 
perturbations. Firstly, this method can only be used after an early
adaptation phase in which the shock adjusts to the changing upstream conditions.
Secondly, this approach is only valid for a C-type shock interacting with a 
density perturbation. For a J-type shock, a second neutral subshock forms 
in the precursor. Behind this secondary shock, the velocities of the neutral and
charged fluids are not the same. Such a situation, which is likely to occur
in a clumpy medium, cannot be modelled with a quasi-steady approach.  

In quiescent cold clouds, silicon is nearly depleted from the gas phase 
and stored in the core of dust grains. However, observations of SiO 
emission near protostellar objects show that some silicon is returned 
to the gas-phase.
It is thought that this is due to sputtering of the dust grains and 
grain-grain collisions in shocks \citep{MPal92}. Although these processes
are not included in the current code, the change of the grain-neutral
relative velocity serves as an indication. 
Grain mantle sputtering requires a grain-neutral relative velocity of 
$\approx$ 6 km s$^{-1}$ whereas core sputtering requires a much higher 
relative velocity of $\approx$ 19 km s$^{-1}$ \citep{Cal97}. 
For our model parameters, with a shock speed of 25~km\ s$^{-1}$, the threshold
for core sputtering is only just attainable. Therefore, we do not expect
much SiO emission from these models. However, grain mantle sputtering 
occurs in all the initial C-type shocks. 
\cite{Jal08} have suggested that if the grain mantle consists of a small 
fraction of silicon, sputtering by heavy molecules such as CO could erode 
the mantles even in low velocity shocks and account for the narrow SiO 
line emission observed in some young outflows \citep[e.g. L1448-mm][]{Jal04}. 
Silicon erosion from the mantle would be saturated for shock velocities 
between 10 and 20 km s$^{-1}$. This implies that the SiO enhancement 
from mantle erosion is saturated in all our models. Variations of the SiO
emission then arise as the spatial area over which sputtering occurs changes
(i.e. the shock width increases/decreases). Furthermore, we expect a small
contribution to mantle sputtering from reflected waves.
In subsequent papers we will include grain mantle and core sputtering as 
well as grain-grain collision terms to describe the release of SiO 
from the grains to the gas phase and use this to calculate the SiO emission
\citep{Cal97,Jal08}.

\section*{Acknowledgements}
We thank the referee, Pierre Lesaffre, for his constructive comments
that improved the paper. The authors thank STFC for the financial support.

\label{lastpage}
\end{document}